\documentclass[review,number]{elsarticle}

\PassOptionsToPackage{hyphens}{url}
\usepackage{hyperref}

\usepackage{latexsym} 
\usepackage{graphics}
\usepackage{epsfig}
\usepackage{graphicx}
\usepackage{tikz-qtree}
\usepackage{tikz}
\usetikzlibrary{shapes, arrows.meta, positioning}
\usepackage{amsmath}
\usepackage{dirtytalk} 
\usepackage{bbm} 
\usepackage{algorithm}
\usepackage{algpseudocode}
\usepackage{pgfplots}
\usepgfplotslibrary{groupplots}
\usepackage[utf8]{inputenc}
\usepackage[english]{babel}
\usepackage{graphicx}
\usepackage{subcaption}
\usepackage{tikz}
\usepackage{graphicx}
\usepackage{subcaption}
\usepackage{pgfplots}
\usepackage{changes}
\pgfplotsset{width=11cm,compat=newest}
\usepackage[moderate]{savetrees} 
\usetikzlibrary{positioning, shapes.geometric}
\usepackage{float} 
\usepackage{colortbl} 
\usepackage{algorithm}
\usepackage{booktabs}
\makeatletter
\def\BState{\State\hskip-\ALG@thistlm}
\makeatother
\algdef{SE}[DOWHILE]{Do}{doWhile}{\algorithmicdo}[1]{\algorithmicwhile\ #1}%

\usepackage{varwidth}
\usepackage{booktabs} 
\usepackage{epsfig}
\usepackage{amsfonts}
\usepackage{array}

\usepackage{rotating}
\usepackage{tabularx}
\usepackage{color}
\usepackage{multirow}
\usepackage{url}
\usepackage{lscape}
\usepackage{xargs} 
\newcolumntype{x}[1]{>{\centering\let\newline\\\arraybackslash\hspace{0pt}}p{#1}}
\usepackage{pgfplots}
\pgfplotsset{compat=1.17}

\usepackage{tabularx}
\usepackage{array}
\usepackage{colortbl}
\usepackage{multirow}
\usepackage{datetime}
\usepackage[mathscr]{eucal}
\usepackage{amssymb}
\usepackage{amsmath}
\usepackage{graphicx}
\usepackage{derivative}
\usepackage[utf8]{inputenc}
\usepackage[T1]{fontenc}
\usepackage{mathtools}
\usepackage[thinc]{esdiff}
\usepackage{wrapfig}
\usepackage{caption}
\usepackage{lscape}
\usepackage{rotating}
\usepackage{url}
\usepackage{babel}
\usepackage{translator}
\usepackage{pgfkeys,pgfcalendar}
\usepackage{graphicx}
\usepackage{multirow}
\usepackage{hhline}
\usepackage{diagbox}
\usepackage{eurosym}
\usepackage{amssymb}
\usepackage{pifont}
\graphicspath{ {/home/ciaran/Downloads/} }
\usepackage{acronym}
\usepackage{booktabs}
\def\preparecolorrefs#1{%
  \setcounter{refindex}{0}%
  \whiledo{\value{refindex}<#1}{%
    \stepcounter{refindex}%
    \expandafter\def\csname\therefindex color\endcsname{black}%
  }%
}
\usepackage{hyperref}

\journal{Applied Energy}

\begin{document}

\begin{frontmatter}

\title{The Evolution of Probabilistic Price Forecasting Techniques: A Review of the Day-Ahead, Intra-Day, and Balancing Markets}

\author{Ciaran O'Connor}
\address{SFI CRT in Artificial Intelligence, School of Computer Science \& IT, University College Cork, Ireland}
\ead{119226305@umail.ucc.ie}

\author{Mohamed Bahloul}
\address{Water \& Energy Transition Unit, Vlaamse Instelling voor Technologisch Onderzoek, Mol, Belgium}
\ead{mohamed.bahloul@vito.be}

\author{Steven Prestwich, Andrea Visentin}
\address{SFI Insight Centre for Data Analytics, School of Computer Science \& IT, University College Cork, Ireland}
\ead{s.prestwich@cs.ucc.ie, andrea.visentin@ucc.ie}

\begin{abstract}
Electricity price forecasting has become a critical tool for decision-making in energy markets, particularly as the increasing penetration of renewable energy introduces greater volatility and uncertainty. Historically, research in this field has been dominated by point forecasting methods, which provide single-value predictions but fail to quantify uncertainty. However, as power markets evolve due to renewable integration, smart grids, and regulatory changes, the need for probabilistic forecasting has become more pronounced, offering a more comprehensive approach to risk assessment and market participation.
This paper presents a review of probabilistic forecasting methods, tracing their evolution from Bayesian and distribution based approaches, through quantile regression techniques, to recent developments in conformal prediction. Particular emphasis is placed on advancements in probabilistic forecasting, including validity-focused methods which address key limitations in uncertainty estimation. Additionally, this review extends beyond the Day-Ahead Market to include the Intra-Day and Balancing Markets, where forecasting challenges are intensified by higher temporal granularity and real-time operational constraints. We examine state of the art methodologies, key evaluation metrics, and ongoing challenges, such as forecast validity, model selection, and the absence of standardised benchmarks, providing researchers and practitioners with a comprehensive and timely resource for navigating the complexities of modern electricity markets.
\end{abstract}


\end{frontmatter}

\section{Introduction}\label{sec:introduction}
Electricity Price Forecasting (EPF) plays a pivotal role in energy markets, enabling market participants to optimise trading strategies, mitigate financial risks, and maintain grid stability. However, forecasting accuracy has become increasingly difficult due to the rapid expansion of renewable energy sources such as wind, solar, and hydroelectric power. While policy incentives, declining technology costs, and regulatory mechanisms like feed-in tariffs have accelerated renewable adoption \cite{koecklin2021public, meles2022adoption}, these energy sources introduce significant variability and uncertainty into electricity markets. Unlike traditional commodities, electricity cannot be stored efficiently, requiring real-time balancing of supply and demand. As a result, short-term price fluctuations are highly sensitive to renewable generation patterns, regulatory interventions, and unforeseen disruptions such as generator outages or transmission constraints. These complexities make EPF particularly challenging, as prices exhibit high volatility, non-linearity, and sudden spikes. Addressing these challenges requires robust forecasting methods capable of quantifying uncertainty, an area that has gained increasing attention in recent years.

\subsection{Probabilistic Forecasting Methods}
There are two major EPF schemes: point forecasts and probabilistic forecasts. Point forecasts provide single price predictions that are easy to interpret and establish parsimonious predictor-target relationships under the assumption of homoscedasticity. While point forecasting has long been the dominant approach in EPF, its inability to quantify uncertainty associated with predictions limits its effectiveness in volatile electricity markets. Price fluctuations driven by renewable energy variability, sudden demand shifts, and regulatory interventions create forecasting challenges that traditional point estimates fail to capture. As a result, probabilistic forecasting has emerged as a key advancement, offering a way to quantify uncertainty by generating Prediction Intervals (PIs) that provide a range of possible future prices rather than a single deterministic value. By capturing the full spectrum of potential price variations, Probabilistic Electricity Price Forecasting (PEPF) enables improved decision-making in dynamic, non-linear markets \cite{nowotarski2018recent, ziel2018probabilistic}, addressing uncertainties in smart grids, supply-demand dynamics, and price fluctuations with a stronger emphasis on the operational impact of forecasts and risk management \cite{khajeh2022applications}.

A variety of probabilistic forecasting methods have been explored in the literature. Traditional approaches include Bayesian models, Historical Simulation, bootstrapped PIs, and Quantile Regression (QR), all of which estimate probability distributions or quantiles of future prices. These methods have been widely applied in electricity markets \cite{nowotarski2018recent, ziel2018probabilistic} but often suffer from limitations in non-stationary environments, where price dynamics are constantly evolving. Since GEFCom2014, Quantile Regression Averaging (QRA) has gained popularity for its ability to combine multiple probabilistic forecasts, improving robustness in complex market conditions \cite{khajeh2022applications}. Despite these advancements, many probabilistic forecasting methods fail to provide coverage guarantees, particularly in data-limited settings. Conformal Prediction (CP) has gained increasing attention for its ability to produce valid and adaptive PIs, ensuring predefined confidence levels are met regardless of the underlying data distribution \cite{gammerman1998learning, vovk2005algorithmic}. Recent advancements, such as Ensemble Batch Prediction Intervals (EnbPI) \cite{xu2021conformal} and Sequential Predictive Conformal Inference (SPCI) \cite{xu2023sequential}, have extended CP to time-series applications, more effectively addressing forecasting challenges in dynamic markets.

\subsection{Related Work and Literature Gap}
EPF has been widely studied, with early research primarily focused on point forecasting techniques. Foundational reviews, such as those by \cite{aggarwal2009electricity, zhang2019review, lago2021forecasting, acarouglu2021comprehensive, jkedrzejewski2022electricity, o2024electricity, cmsf2025011016, o2025review}, catalogued the development of statistical, ML, DL, and hybrid models best suited for EPF. However, these reviews emphasise methodological innovation while largely overlooking the growing need to quantify uncertainty in increasingly volatile electricity markets. In contrast, probabilistic forecasting has received far less attention in EPF literature. The few existing reviews that address uncertainty quantification, such as those by \cite{weron2014electricity} and \cite{nowotarski2018recent}, primarily focus on traditional techniques and remain largely confined to the Day-Ahead Market (DAM). They do not incorporate recent methodological advancements, particularly validity-focused approaches such as CP and its adaptations for time series forecasting.

Another important gap in the literature is the under-representation of real-time electricity spot markets. While the DAM has been extensively studied, the Intra-Day Market (IDM) and Balancing Market (BM) remain largely unexplored, despite their increasing relevance in renewable-heavy systems. These markets introduce unique forecasting challenges due to higher temporal granularity, greater price volatility, and shorter decision horizons. Addressing these complexities is critical for effective market participation, grid reliability, and operational risk management. Given the rapid evolution of probabilistic forecasting techniques and the growing importance of real-time electricity markets, there is a clear need for a comprehensive and up to date review. This paper addresses this gap by examining probabilistic forecasting methods across the DAM, IDM, and BM. We trace the methodological progression from parametric and regression based models to recent CP-based approaches that offer rigorous validity guarantees. In doing so, this review provides an up-to-date resource for researchers and industry practitioners navigating the complexities of EPF.

The structure of this paper is as follows. Section \ref{Market_Structure_Review} provides a background on electricity spot markets, including the DAM, IDM, and BM. Section \ref{Uncertainty_Review} reviews uncertainty quantification methods used in PEPF. In Section \ref{Discussion}, we discuss the key findings for each market. Finally, Section \ref{conclusionsec} summarises the contributions and outlines future research opportunities in probabilistic electricity price forecasting.

\section{Background: Electricity Market Structure}\label{Market_Structure_Review}  
\begin{figure}[ht!]
\centering
\begin{tikzpicture}[scale=0.89] 

\tikzstyle{market} = [ellipse, draw, thick, minimum width=2.35cm, minimum height=1.9cm, text centered, font=\bfseries\small, align=center]
\tikzstyle{financial} = [market, color=cyan, fill=cyan!10]
\tikzstyle{physical} = [market, color=blue, fill=blue!20]
\tikzstyle{physical1} = [market, color=blue, fill=blue!16]
\tikzstyle{physical2} = [market, color=blue, fill=blue!12]
\tikzstyle{physical3} = [market, color=blue, fill=blue!8]
\tikzstyle{separator} = [thick, draw=gray]

\draw[separator] (-5.6, 1.5) -- (-3.1, 1.5); 
\draw[separator] (-2.6, 1.5) -- (7.6, 1.5);   

\node at (-4.2, 1.8) {\footnotesize \textbf{Financial}};
\node at (2.4, 1.8) {\footnotesize \textbf{Physical}};

\node[financial] (forwards) at (-4.2, 0) {\footnotesize Forwards/ \\ Capacity/ \\ FTR};

\node[physical] (dayahead) at (-1.15, 0) {\scriptsize Day-Ahead};
\node[physical1] (intraday) at (1.6, 0) {\scriptsize Intraday};
\node[physical2] (balancing) at (4.2, 0) {\scriptsize Balancing};
\node[physical3] (dispatch) at (6.8, 0) {\scriptsize Dispatch};

\draw[separator] (-5.6, -1.5) -- (7.6, -1.5);
\node at (-4.8, -1.8) {\tiny Years};
\node at (-3.2, -1.8) {\tiny Months};
\node at (-1.6, -1.8) {\tiny Weeks};
\node at (0.2, -1.8) {\tiny Days};
\node at (2.2, -1.8) {\tiny Hours};
\node at (4.2, -1.8) {\tiny Minutes};
\node at (5.9, -1.8) {\tiny Seconds};
\node at (7.2, -1.8) {\tiny Real-time};

\end{tikzpicture}
\caption{Contrasting granularity of markets}
\label{Market_Outline_elec}
\end{figure}
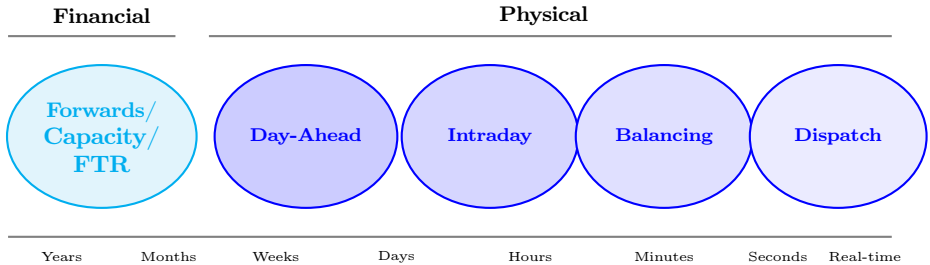

EPF is closely linked to market structure, as different electricity markets operate on distinct time frames (see Figure \ref{Market_Outline_elec}), settlement mechanisms, and exhibit varying levels of temporal granularity and price volatility. Short-term markets, including the DAM, IDM, and BM, are essential for maintaining real-time balance between supply and demand. Each presents unique forecasting challenges due to differences in trading horizons, data availability, volatility, and operational constraints. Understanding the structure and role of each market is essential for selecting appropriate forecasting methods tailored to their specific characteristics.

\subsection{Day-Ahead Market}\label{Backgroun_DAM}
The DAM is a forward electricity market in which trades are scheduled one day in advance, typically in hourly blocks. Market participants submit bids and offers based on anticipated supply and demand, with market clearing conducted via optimisation algorithms such as EUPHEMIA in European markets \cite{newbery2016benefits, ilea2018european}. While the DAM generally exhibits lower price volatility than real-time markets, forecasting remains challenging due to the increasing penetration of weather-dependent renewables. Wind and solar generation introduce forecast error and ramping uncertainty, complicating both supply and demand projections \cite{martinez2016impact}. The DAM’s longer forecast horizon and lower volatility have historically favoured point forecasting methods, but the integration of intermittent generation is driving greater interest in uncertainty-aware approaches.

\subsection{Intra-Day Market}\label{Backgroun_IDM}
The IDM enables continuous trading in the hours leading up to physical delivery, allowing market participants to adjust their DAM positions in response to updated forecasts, asset availability, and real-time system conditions. Unlike the DAM, which clears once daily, the IDM features multiple gate closures, often 15 to 60 minutes ahead of delivery, making it a short lead-time, high-resolution market. The IDM is highly sensitive to short-term deviations in renewable output, outages, and intra-day demand shifts \cite{shinde2019literature}. Accurate and timely forecasts are therefore critical for imbalance mitigation, trading optimisation, and system reliability. The non-stationary nature of IDM prices and their frequent fluctuations limit the effectiveness of point forecasting, necessitating probabilistic models that can adapt to dynamic and uncertain conditions.

\subsection{Balancing Market}\label{Backgroun_BM}
The BM is a real-time market operated by the Transmission System Operator (TSO) to correct supply-demand imbalances that remain after IDM gate closure. Unlike other markets, the BM does not involve voluntary trading; instead, the TSO dispatches balancing energy based on real-time system needs, typically with settlement periods as short as 5 minutes \cite{ortner2019future, zachmann2023design}. Price formation in the BM reflects the marginal cost of these last-minute corrections, leading to extreme volatility and frequent price spikes \cite{dumas2019probabilistic}. Forecasting imbalance prices is especially difficult due to limited transparency, highly non-linear dynamics, and the need for fast reaction times. In this context, probabilistic forecasting methods are essential for quantifying risk and preparing for worst-case outcomes. Integrating real-time signals such as frequency deviations, forecast errors, and system constraints can significantly enhance forecasting accuracy and decision-making in the BM.

\subsection{Other Electricity Markets}
In addition to the DAM, IDM, and BM, several other electricity markets support long-term planning, system reliability, and financial hedging. Below is a brief overview of their roles and relevance from a forecasting perspective:
\begin{itemize}
    \item \textit{Forward Market}: Enables market participants to hedge future positions in the DAM, IDM, and BM through contracts-for-difference \cite{peura2021renewable}. These contracts lock in a strike price for a future delivery period, reducing exposure to spot price volatility. Forecasting in this context involves long-term price trend analysis, typically using macroeconomic indicators, fuel price forecasts, and policy-driven demand models—distinct from the short-term, high-resolution focus of this review.

    \item \textit{Ancillary Markets}: Maintain grid stability by procuring services such as frequency regulation, spinning reserve, voltage support, and black start capabilities. Forecasting in these markets involves real-time operational signals (e.g., frequency deviations, ramping needs) and often relies on probabilistic or control-oriented models, rather than pure price forecasting \cite{rancilio2022ancillary}.

    \item \textit{Capacity Market}: Ensures system adequacy by compensating providers for committing firm capacity during peak periods. Participation is typically based on long-term commitments, and forecasting focuses on capacity availability and peak demand projections across seasonal or annual horizons \cite{cramton2013capacity}. These differ significantly from the short-term volatility and data cadence addressed in this review.

    \item \textit{Financial Transmission Rights Auctions}: Provide financial hedges against locational price differences caused by transmission congestion. FTRs entitle holders to payments based on the spread between nodal prices across network zones. Forecasting in this domain requires congestion modelling and transmission flow forecasting, often using power flow simulations and market coupling dynamics \cite{sarkar2008comprehensive}.
\end{itemize}
While forward, ancillary, capacity, and transmission rights markets are critical to the broader functioning of electricity systems, they fall outside the scope of this review. These markets operate on longer timescales, involve structural risk hedging or reliability assurance, and often require forecasting methodologies distinct from those used for short-term spot price prediction. This paper therefore focuses exclusively on the DAM, IDM, and BM, where short-term price volatility, renewable integration, and real-time trading decisions require probabilistic forecasts.

\section{Uncertainty Estimation in Electricity Price Forecasting}\label{Uncertainty_Review}
Uncertainty in EPF can be broadly classified into two categories: epistemic uncertainty, arising from limited knowledge or data, and aleatoric uncertainty, reflecting inherent randomness in the system. This review focuses on aleatoric uncertainty, which, although irreducible, can be modelled probabilistically \cite{kendall2017uncertainties, der2009aleatory, hacking2006emergence, sullivan2015introduction, hullermeier2021aleatoric}.
Thus, for addressing both epistemic and aleatoric uncertainty we look at various probabilistic forecasting methods, including Bayesian, Distribution Based, Monte Carlo, Bootstrap, Historical Simulation, QR \cite{koenker1978regression}, QRA \cite{nowotarski2015computing}, CP \cite{gammerman1998learning}, EnbPI \cite{xu2021conformal}, SPCI \cite{xu2023sequential}, and Conformalised QR. All these approaches have received increasing attention for capturing uncertainties in energy markets as well as various other time series applications \cite{nowotarski2018recent, ziel2018probabilistic}. Recent innovations have focused on offering more reliable PIs, including making CP methods applicable to time series data. This section reviews these techniques, emphasising their role in addressing market variability and uncertainty.

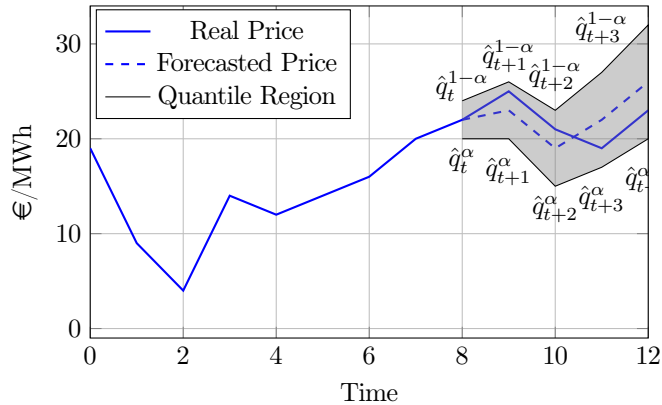
\begin{figure}[ht]
    \centering
    \begin{tikzpicture}
        \begin{axis}[
            width=9cm, 
            height=6cm,
            xlabel={Time},
            ylabel={\euro /MWh},
            title={},
            grid=both,
            xmin=0, xmax=12, 
            ymin=-1, ymax=34, 
            legend style={at={(0.01,0.99)}, anchor=north west}
        ]

        \addplot[blue, thick] coordinates {
            (0, 19) (1, 9) (2, 4) (3, 14) (4, 12) (5, 14) (6, 16) (7, 20) (8, 22) (9, 25) (10, 21) (11, 19) (12, 23)
        } ;
        \addlegendentry{Real Price}
        
        \addplot[blue, dashed, thick] coordinates {
            (8, 22) (9, 23) (10, 19) (11, 22) (12, 26)
        };
        \addlegendentry{Forecasted Price}
        
        \addplot[fill=gray, fill opacity=0.4] coordinates {
            (8, 20) (9, 20) (10, 15) (11, 17) (12, 20) 
            (12, 32) (11, 27) (10, 23) (9, 26) (8, 24) (8, 24)
        };
        \addlegendentry{Quantile Region}
        
        \node at (axis cs:8,20) [anchor=north] {$\hat{q}^{\alpha}_{t}$};
        \node at (axis cs:8,23) [anchor=south] {$\hat{q}^{1-\alpha}_{t}$};
        
        \node at (axis cs:9,19) [anchor=north] {$\hat{q}^{\alpha}_{t+1}$};
        \node at (axis cs:9,26) [anchor=south] {$\hat{q}^{1-\alpha}_{t+1}$};
        
        \node at (axis cs:10,15) [anchor=north] {$\hat{q}^{\alpha}_{t+2}$};
        \node at (axis cs:10,24) [anchor=south] {$\hat{q}^{1-\alpha}_{t+2}$};
        \node at (axis cs:11,16) [anchor=north] {$\hat{q}^{\alpha}_{t+3}$};
        \node at (axis cs:11,29) [anchor=south] {$\hat{q}^{1-\alpha}_{t+3}$};
        \node at (axis cs:12,18) [anchor=north] {$\hat{q}^{\alpha}_{t+4}$};
        \node at (axis cs:12,-100) [anchor=south] {$\hat{q}^{1-\alpha}_{t+4}$};    
        \end{axis}
    \end{tikzpicture}
    \caption{Quantile forecast of electricity prices}
    \label{QuantilePlot}
\end{figure}

We approach EPF as a regression task, where the dataset \(\mathcal{Z} = \{(\mathbf{x}_t, y_t)\}_{t=1}^N\) consists of predictor variables \( x_t \) and corresponding response variables \( y_t \). Instead of providing single-value predictions, probabilistic forecasting methods estimate a range within which future prices are likely to fall. The conditional quantile function \( Q_{y_t}(\alpha \mid \mathbf{x}_t) \) defines the value below which a proportion \( \alpha \) of future price observations are expected, given the predictors \( x_t \). We denote the lower and upper quantiles at levels \( \alpha \) and \( 1-\alpha \) as \( \hat{q}_t^\alpha \) and \( \hat{q}_t^{1-\alpha} \), respectively. The PI produced by forecasting model \( f \) is defined as:
\begin{equation}
\Gamma_{1-2\alpha}(x_t) = [\hat{q}_t^\alpha, \hat{q}_t^{1-\alpha}],
\end{equation}
quantifying forecast uncertainty by indicating that the future price \( y_t \) is expected to fall within this range with a confidence level of \( 1-2\alpha \), as shown in Figure \ref{QuantilePlot}. For example, setting \( \alpha = 0.1 \) yields an 80\% confidence interval \( (1-2\alpha = 0.8) \), meaning there is an 80\% probability that \( y_t \) will lie within the estimated range.

\subsection{Parametric Methods} 
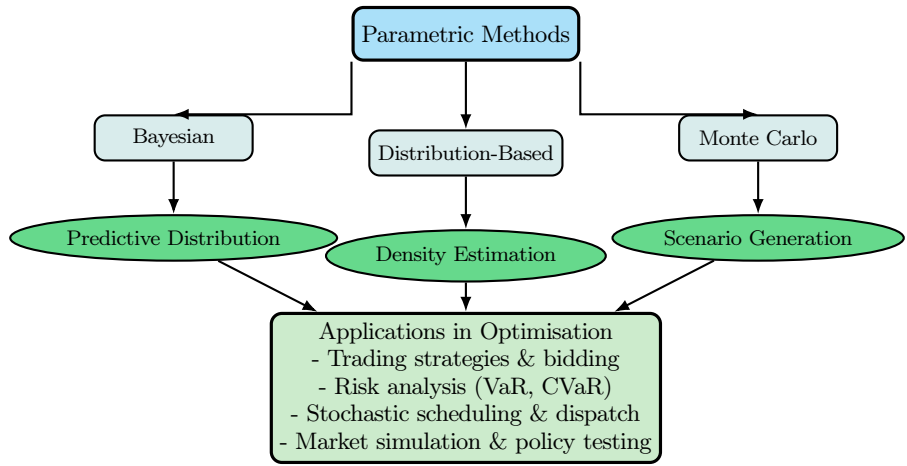
\begin{figure}[ht]
\centering
\begin{tikzpicture}[
  node distance=7mm and 11mm, 
  family/.style={rectangle, rounded corners, draw=black, very thick, fill=cyan!30,
                 minimum width=3.0cm, minimum height=0.7cm, font=\small, align=center},
  method/.style={rectangle, rounded corners, draw=black, thick, fill=teal!15,
                 minimum width=2.1cm, minimum height=0.6cm, font=\footnotesize, align=center},
  output/.style={ellipse, draw=black, thick, fill=green!50!teal!60,
                 minimum width=2.2cm, minimum height=0.6cm, font=\footnotesize, align=center},
  final/.style={rectangle, rounded corners, draw=black, very thick, fill=green!60!black!20,
                minimum width=3.3cm, minimum height=0.8cm, font=\small, align=center},
  arrow/.style={-{Latex[length=2mm]}, thick}
]

\node[family] (parametric) {Parametric Methods};

\node[method, below left=of parametric, xshift=-2mm] (bayesian) {Bayesian};
\node[method, below=of parametric, yshift=-2mm] (distribution) {Distribution-Based};
\node[method, below right=of parametric, xshift=2mm] (montecarlo) {Monte Carlo};

\node[output, below=of bayesian] (full) {Predictive Distribution};
\node[output, below=of distribution] (density) {Density Estimation};
\node[output, below=of montecarlo] (scenarios) {Scenario Generation};

\node[final, below=18mm of distribution] (decision) {Applications in Optimisation
\\- Trading strategies \& bidding
\\- Risk analysis (VaR, CVaR)
\\- Stochastic scheduling \& dispatch
\\- Market simulation \& policy testing};

\draw[arrow] (parametric.south west) |- (bayesian.north);
\draw[arrow] (parametric.south) -- (distribution.north);
\draw[arrow] (parametric.south east) |- (montecarlo.north);

\draw[arrow] (bayesian) -- (full);
\draw[arrow] (distribution) -- (density);
\draw[arrow] (montecarlo) -- (scenarios);

\draw[arrow] (full) -- (decision);
\draw[arrow] (density) -- (decision);
\draw[arrow] (scenarios) -- (decision);

\end{tikzpicture}
\caption{Parametric probabilistic forecasting methods: Bayesian, Distribution-Based, and Monte Carlo.}
\label{ParamPlot}
\end{figure}

Parametric methods assume that electricity price dynamics follow a specific probability distribution (e.g., Gaussian, Gamma) and estimate its parameters from historical data. These approaches facilitate the incorporation of prior knowledge and produce interpretable forecasts; however, their validity depends on the accuracy of the assumed distribution, which can limit their ability to consistently achieve nominal PI coverage. An outline of the methods can be seen in Figure \ref{ParamPlot}.

\subsubsection{Bayesian}
Bayesian probabilistic forecasting provides a flexible framework for uncertainty quantification in PEPF, incorporating prior distributions to model uncertainty and updating beliefs as new data becomes available.  At the core of Bayesian forecasting is the posterior distribution, obtained via Bayes’ theorem:
\begin{equation}
p(\theta | \mathcal{D}) = \frac{p(\mathcal{D} | \theta) p(\theta)}{p(\mathcal{D})}
\end{equation}
where \( p(\theta | \mathcal{D}) \) is the posterior distribution of parameters given data \( \mathcal{D} \), \( p(\mathcal{D} | \theta) \) is the likelihood, \( p(\theta) \) is the prior, and \( p(\mathcal{D}) \) is the marginal likelihood. Bayesian models such as Bayesian Neural Networks (BNNs), Gaussian Processes, and state-space models have been applied to EPF, leveraging Markov Chain Monte Carlo (MCMC), variational inference, and Bayesian model averaging (BMA) for enhanced predictive performance. These methods have demonstrated success in capturing time-varying market dynamics, modelling stochastic volatility, and refining probabilistic interval estimates. 

BNNs and other Bayesian modelling techniques have gained traction in DAM EPF, particularly for their ability to provide full predictive distributions while mitigating overfitting. \cite{vahidinasab2010bayesian} demonstrates the effectiveness of BNNs in DAM forecasting, highlighting their capacity to capture uncertainty in price predictions. Beyond electricity markets, BNNs and other Bayesian methods have also proven effective in cloud workload prediction \cite{rossi2022bayesian, rossi2023clustering, rossi2025forecasting}, underscoring their adaptability in stochastic environments. Further developments in Bayesian inference have improved forecasting reliability by incorporating market-specific characteristics. \cite{muller2019bayesian} explores Bayesian inference for Continuous-time Autoregressive Moving Average (CARMA) models, demonstrating improved probabilistic forecasts through volatility and seasonality modelling. Bayesian optimisation has also been employed to enhance hybrid forecasting frameworks, as shown by \cite{cheng2019hybrid}, where hyperparameter tuning improves predictive accuracy. \cite{kostrzewski2019probabilistic} applies Bayesian stochastic volatility models with jumps, generating more precise PIs. Similarly, \cite{ghayekhloo2019combination} integrates Bayesian learning into Recurrent Neural Networks (RNNs), improving the robustness of probabilistic forecasts. The use of Bayesian methodologies in deep learning (DL) is further demonstrated in \cite{brusaferri2019bayesian}, where Bayesian learning is incorporated into deep neural networks (DNNs) for DAM forecasting. In a comparative study, \cite{brusaferri2020probabilistic} evaluates BNNs against Mixture Density RNNs, illustrating Bayesian inference’s role in risk assessment. Additionally, \cite{bozorg2020bayesian} applies Bayesian bootstrap QR to photovoltaic power forecasting, leveraging resampled data for improved accuracy. BMA has emerged as another promising technique, as discussed by \cite{bhatia2021ensemble}, where it improves forecast reliability by aggregating multiple predictive models. Hybrid Bayesian approaches, which integrate BNNs with Monte Carlo simulation and hierarchical frameworks, have further improved forecasting accuracy by capturing causal relationships and market uncertainties \cite{yuanchen2023electricity, nickelsen2024bayesian}. \cite{yang2025data} extends these methodologies by developing a Bayesian inference-based calibration framework for wholesale electricity price simulation, incorporating Bayesian Ridge Regression for extreme price spikes and Dual-head BNNs for more stable price periods. 

While Bayesian methods have been extensively studied in the DAM, their application in the IDM remains limited. \cite{panagiotelis2008bayesian} introduces a Bayesian framework for IDM price forecasting using multivariate skew t-distributions, effectively capturing key time series features such as skewness and heavy tails. In a separate study, \cite{cottet2003bayesian} develops a Bayesian multi-equation regression model for IDM electricity load forecasting in New South Wales, employing Bayesian model selection and MCMC sampling to enhance predictive distributions and better quantify short-term uncertainty. More recently, \cite{klein2023deep} proposes a deep distributional time series model for probabilistic IDM forecasting in the Australian market, leveraging Bayesian Echo State Networks and Gaussian copula processes to improve uncertainty quantification and tail risk estimation. \cite{nickelsen2025bayesian} provides a comprehensive survey of deep Bayesian forecasting models, demonstrating improved calibration and sharpness in probabilistic predictions through the integration of epistemic and aleatoric uncertainty estimation. These studies suggest that Bayesian methods can be effectively adapted to the IDM, particularly for modelling price fluctuations in continuous trading environments. 

In contrast, Bayesian forecasting remains largely underexplored in the BM, with the only notable application by \cite{lima2022bayesian} demonstrating that while dynamic linear models did not consistently outperform GARCH variants at lower quantiles, their time-varying predictive densities showed superior average accuracy and stronger performance in capturing extreme values, highlighting Bayesian methods as a promising yet underutilised approach for robust PEPF in real-time markets. However, further research is needed to assess their computational feasibility in real-time operations.

\subsubsection{Distribution-Based Probabilistic Forecasts}
Distribution-based probabilistic forecasting aims to model the full probability distribution of electricity prices. Formally, given historical observations \( \mathcal{D} \), distribution-based methods seek to estimate the conditional density function \( p(y \mid \mathbf{x}) \), where \( y \) represents the electricity price and \( \mathbf{x} \) denotes predictor variables. A common nonparametric approach is Kernel Density Estimation (KDE), which approximates the density as:
\begin{equation}
\hat{p}(y \mid \mathbf{x}) = \frac{1}{N h} \sum_{t=1}^{N} K \left(\frac{y - y_t}{h} \right),
\end{equation}
where \( K(\cdot) \) is a kernel function (e.g., Gaussian), and \( h \) is the bandwidth parameter controlling smoothness. Parametric methods, such as Generalized Additive Models for Location, Scale, and Shape (GAMLSS), estimate distribution parameters dynamically:
\begin{equation}
y_t \sim F(\mu_t, \sigma_t, \nu_t, \tau_t),
\end{equation}
where \( F \) represents a flexible parametric family (e.g., Gaussian, Beta, or Gamma), and \( \mu_t, \sigma_t, \nu_t, \tau_t \) are time-varying parameters estimated using penalized likelihood methods. Recent advances integrate DL with distributional modelling, leveraging QR with kernel smoothing, copula-based modelling, and Distributional DNNs (DDNNs) to generate predictive densities that adapt to market volatility. 

In the DAM, nonparametric and semi-parametric techniques have gained attention for their ability to model complex price distributions. \cite{weron2008forecasting} compares parametric and KDE-based models, finding that KDE produces smoother and more reliable probabilistic forecasts, particularly in capturing market fluctuations. Expanding on this, \cite{serinaldi2011distributional} introduces GAMLSS, dynamically modelling price distributions to generate flexible and well-calibrated PIs. Beyond standard time series modelling, \cite{ziel2018probabilistic} extends the X-Model by incorporating supply and demand curves, simulating time-dependent distributions that capture extreme price events, offering improved risk assessment capabilities. \cite{dudek2018probabilistic} proposes a Nadaraya-Watson estimator, showing that its asymmetric predictive densities better capture price spikes than traditional models. Similarly, \cite{monteiro2018new} applies KDE to estimate Beta PDFs, providing a comprehensive uncertainty profile for DAM prices. Further refining these techniques, \cite{klein2023deep} integrates echo state networks with Gaussian copula processes, generating well-calibrated probabilistic forecasts even in highly volatile price regimes. DL and hybrid models have further advanced distribution-based forecasting. \cite{marcjasz2023distributional} introduces DDNNs, modelling full electricity price distributions to improve risk assessment and forecasting accuracy. \cite{xu2024novel} combines QR and Long Short-Term Memory with KDE, effectively converting quantile forecasts into full probabilistic distributions. Lastly, \cite{lipiecki2024postprocessing} employs Isotonic Distributional Regression as a postprocessing technique, refining probabilistic forecasts and outperforming traditional methods such as QRA and CP in volatile market conditions.

Despite their extensive use in the DAM, distribution-based methods remain underexplored in the IDM, where capturing full price distributions is crucial for market participants navigating its higher volatility and continuous trading structure. \cite{cramer2023multivariate} leverages normalising flows to generate distributional forecasts for IDM prices, effectively modelling the price spread between DAM and IDM as a multivariate probability distribution. \cite{klein2023deep} extends their DAM work by applying deep distributional time series models in the Australian IDM, incorporating probabilistic demand predictions to improve upper-tail accuracy—an essential factor given the IDM's heightened price fluctuations. Additionally, \cite{hirsch2024simulation} explores a simulation-based probabilistic forecasting model, demonstrating that distribution-based forecasts substantially enhance tail risk estimation and volatility modelling, particularly in IDM settings where forecast updates, outages, and the merit-order effect drive price uncertainty. \cite{chen2025probabilistic} unify marginal and multivariate dependency modelling in a single framework using a scoring rule-based generative approach to forecast full price paths in the German intraday market, evaluating both statistical accuracy and economic value in trading strategies.

Among all short-term electricity markets, the BM has received the least attention in the context of distribution-based forecasting. The only notable application is found in \cite{narajewski2022probabilistic}, where probabilistic forecasting methods, including GAMLSS and probabilistic neural networks (NNs), are applied to the German BM. While these models improve empirical coverage, the study finds that they do not significantly outperform simple intraday continuous price indices in predictive accuracy. This suggests that, while distribution-based models offer theoretical advantages in uncertainty quantification, their practical benefit in real-time markets remains uncertain, likely due to the BM's extreme volatility and the difficulty of modelling rapid system imbalances.

\subsubsection{Monte Carlo}
Monte Carlo methods provide a powerful framework for probabilistic forecasting by simulating multiple possible future price trajectories. These methods approximate the probability distribution of electricity prices by generating a large number of random samples from an assumed or estimated stochastic process. Formally, given a probabilistic model for electricity prices \( y_t \) conditioned on predictor variables \( \mathbf{x}_t \), Monte Carlo methods approximate the expected forecast distribution by simulating \( M \) independent realisations:
\begin{equation}
y_t^{(m)} \sim p(y_t \mid \mathbf{x}_t), \quad m = 1, \dots, M.
\end{equation}
The empirical forecast distribution is then obtained as:
\begin{equation}
\hat{p}(y_t \mid \mathbf{x}_t) = \frac{1}{M} \sum_{m=1}^{M} \delta(y_t - y_t^{(m)}),
\end{equation}
where \( \delta(\cdot) \) denotes the Dirac delta function, representing the probability mass at each simulated outcome.  
Monte Carlo methods can be applied across various probabilistic models, including MCMC for Bayesian inference, Sequential Monte Carlo for real-time updates, and Monte Carlo Dropout in DL for uncertainty estimation.

In the DAM, Monte Carlo methods are frequently integrated with time series and econometric models to improve probabilistic forecasts. \cite{wu2010hybrid} employs Monte Carlo simulations alongside ARMAX, GARCH, and Artificial Wavelet NNs, demonstrating that variance reduction techniques enhance forecasting accuracy. \cite{bello2016probabilistic} extends this approach by incorporating Monte Carlo simulations into demand, fuel price, and renewable generation forecasting, generating multiple scenarios that refine medium-term price predictions. Similarly, \cite{bello2016medium} uses Monte Carlo-based risk factor simulations to construct a probabilistic risk profile for medium-term electricity prices, offering insights into uncertainty-driven market fluctuations. Monte Carlo methods extend beyond traditional econometric models into machine learning (ML) and NNs, where they enhance demand forecasting reliability by adjusting weights through random sampling \cite{yong2017neural}. \cite{tat2018electricity} utilises a mean-reverting Monte Carlo model to simulate price trends and spikes in Lithuania’s electricity market, effectively capturing short-term volatility. In hybrid modelling approaches, \cite{osorio2018hybrid} combines Monte Carlo simulations with deterministic price models to generate multiple forecasted price paths, improving uncertainty management in short-term trading strategies. Monte Carlo methods optimize DAM bidding strategies by simulating multiple price trajectories, enabling market participants to refine decisions and develop risk-adjusted trading strategies under uncertainty \cite{narajewski2022optimal}. In a related study, \cite{yildirim2022supply} applies Sequential MCMC techniques to forecast supply curves, enhancing adaptability to real-time price variations and improving volatility estimation. Meanwhile, \cite{yuanchen2023electricity} integrates Bayesian networks with Monte Carlo simulation to capture interdependencies in electricity price distributions, demonstrating improved forecast reliability in volatile DAM conditions. \cite{pavirani2025predicting} applies a Monte Carlo simulation framework to PEPF forecasting in Italy, demonstrating improved capture of extreme price fluctuations by incorporating stochastic renewables, demand uncertainty, and scenario-based system constraints.

Despite their extensive use in DAM forecasting, Monte Carlo methods remain underdeveloped in real-time markets like the IDM, where dynamic probabilistic forecasting is crucial for capturing intraday fluctuations, with \cite{panagiotelis2008bayesian} being a notable example using MCMC for Bayesian density forecasting of IDM prices. MCMC-based inference effectively captures key time series properties like serial correlation, skewness, and heavy tails, improving probabilistic density estimation, but its 30-day testing period leaves its long-term IDM applicability uncertain, whereas the BM has seen more direct Monte Carlo applications due to its extreme volatility and need for scenario-based forecasting. 

\cite{brolin2010modeling} uses Monte Carlo simulations to generate scenario trees that capture BM volatility and DAM price deviations, enhancing decision-support tools and helping market participants anticipate imbalance price fluctuations. More recently, \cite{pavirani2024predicting} applies Monte Carlo Tree Search to BM EPF, achieving a 20.4\% accuracy improvement under ideal conditions and 12.8\% in realistic settings by incorporating system dynamics and implicit demand responses, highlighting the potential of Monte Carlo methods when integrated with optimization-based approaches. Monte Carlo simulations effectively model uncertainty in the DAM for bidding and risk management but remain underexplored in real-time markets. While MCMC shows promise in the IDM, limited testing calls for further validation, whereas the BM has seen greater adoption in scenario-based forecasting and market optimisation.

\subsection{Non-Parametric Methods}  
\begin{figure}[ht]
\centering
\begin{tikzpicture}[
  node distance=6mm and 7mm, 
  family/.style={rectangle, rounded corners, draw=black, very thick, fill=red!30,
                 minimum width=3.6cm, minimum height=0.7cm, font=\small, align=center},
  method/.style={rectangle, rounded corners, draw=black, thick, fill=orange!65,
                 minimum width=2.3cm, minimum height=0.6cm, font=\footnotesize, align=center},
  output/.style={ellipse, draw=black, thick, fill=yellow!70!orange!40,
                 minimum width=2.2cm, minimum height=0.6cm, font=\footnotesize, align=center},
  final/.style={rectangle, rounded corners, draw=black, very thick, fill=yellow!60!red!20,
                minimum width=3.5cm, minimum height=0.8cm, font=\small, align=center},
  arrow/.style={-{Latex[length=2mm]}, thick}
]

\node[family] (nonparam) {Non-Parametric Methods};

\node[method, below left=of nonparam, xshift=-1mm] (bootstrap) {Bootstrap};
\node[method, below right=of nonparam, xshift=1mm] (historical) {Historical Simulation};

\node[output, below=of bootstrap] (empirical) {Empirical Intervals};
\node[output, below=of historical] (residual) {Residual Scenarios};

\node[final, below=28mm of nonparam] (decision) {Applications in Optimisation
\\- Empirical risk calibration
\\- Robust hedging \& stress testing
\\- Intraday bidding \& adjustments
\\- Fast benchmarking \& validation
};

\draw[arrow] (nonparam.south west) |- (bootstrap.north);
\draw[arrow] (nonparam.south east) |- (historical.north);

\draw[arrow] (bootstrap) -- (empirical);
\draw[arrow] (historical) -- (residual);

\draw[arrow] (empirical.south) |- (decision.north);
\draw[arrow] (residual.south) |- (decision.north);

\end{tikzpicture}
\caption{Non-parametric probabilistic forecasting methods: Bootstrap and Historical Simulation.}
\label{NonParamPlot}
\end{figure}
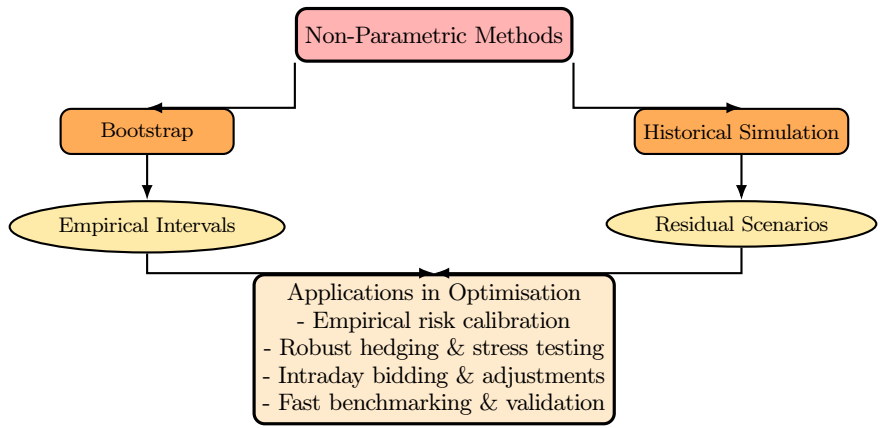

Non-parametric methods do not assume a specific distribution for the data. These methods, like bootstrap and historical simulation, provide flexibility and robustness in environments where data characteristics are unknown or complex. An outline of the methods can be seen in Figure \ref{NonParamPlot}.

\subsubsection{Bootstrap}
Bootstrap methods provide a nonparametric approach to probabilistic forecasting by resampling observed data to approximate the sampling distribution of electricity price forecasts. Unlike traditional parametric models, bootstrap techniques require no explicit distributional assumptions, making them well-suited for capturing forecast uncertainty in volatile electricity markets. Given a historical dataset \( \mathcal{D} \), bootstrap methods generate probabilistic forecasts by drawing \( B \) bootstrap samples \( \mathcal{D}^{(b)} \), where each sample is obtained by resampling with replacement from \( \mathcal{D} \). For each bootstrap sample, a predictive model is trained, producing a set of forecasts \( \{ \hat{y}_t^{(b)} \}_{b=1}^{B} \). The final predictive density is then estimated as:
\begin{equation}
\hat{p}(y_t \mid \mathbf{x}_t) = \frac{1}{B} \sum_{b=1}^{B} \delta(y_t - \hat{y}_t^{(b)}),
\end{equation}
where \( \delta(\cdot) \) represents the Dirac delta function, placing probability mass at each bootstrapped prediction. Variants of the bootstrap, such as residual-based bootstrapping, apply resampling to forecast residuals to improve PIs, while bagging reduces variance in ensemble models by averaging multiple bootstrap-based forecasts. Bayesian bootstrap extends this approach by assigning random weights to resampled observations, improving robustness in Bayesian probabilistic forecasting. 

In the DAM, bootstrap techniques have been integrated into both traditional statistical models and ML frameworks to improve PEPF accuracy. \cite{clements2007bootstrap} introduces bias-corrected bootstrap methods for AR models, effectively addressing small-sample biases and refining forecast PIs. Similarly, \cite{alonso2011seasonal} applies a bootstrap approach to seasonal dynamic factor models, demonstrating improved probabilistic forecasting in electricity markets by resampling residuals. ML-based approaches have further extended the use of bootstrap methods, particularly in hybrid and ensemble models. Bootstrap-integrated Extreme Learning Machines (ELM) have been widely applied to DAM EPF, with \cite{chen2012electricity} and \cite{wan2013hybrid} showing significant improvements in both point forecasts and PIs in the Australian market. More recently, \cite{loizidis2024electricity} demonstrates the effectiveness of bootstrap-ELM in capturing both normal and extreme price fluctuations in the German and Finnish markets. Other NN-based approaches have also benefited from bootstrap enhancements, such as \cite{khosravi2013quantifying}, which constructs PIs in NNs using bootstrap resampling. While this method produces narrower, more informative intervals, it occasionally underperforms in highly volatile conditions due to underestimated uncertainty. Beyond NNs, residual-based bootstrapping improves PEPF accuracy by resampling forecast residuals, with \cite{uniejewski2019importance} demonstrating its superiority over traditional confidence interval methods in providing more reliable uncertainty quantification. Bayesian adaptations of bootstrap methods have also been explored, with \cite{bozorg2020bayesian} applying Bayesian bootstrap QR to improve probabilistic photovoltaic power forecasts, improving quantile estimation through resampled data. Bootstrap Aggregation (Bagging) enhances DAM EPF by reducing forecast variance and capturing non-linear dependencies, with \cite{harasheh2016forecasting} showing its consistent performance improvement over NNs and statistical models in the Italian market, while \cite{bhatia2021ensemble} demonstrates its role in stabilizing ensemble models for EPF applications. A comparative analysis by \cite{zhang2023probabilistic} evaluates bootstrap methods against QR, concluding that while both effectively capture non-linear dependencies, QR provides sharper PIs, suggesting that bootstrap methods may be better suited for capturing broader market uncertainties rather than fine-tuning interval sharpness. 

Although bootstrap applications in the IDM are scarce, studies suggest they balance computational efficiency with forecasting accuracy in continuous trading, as \cite{narajewski2022optimal} demonstrates by using bootstrapped errors to simulate price trajectories in electricity auctions across DAM and IDM, leading to improved profits in both markets. In a related study, \cite{serafin2022trading} applies a bootstrap-based approach to construct prediction bands for intraday electricity price path forecasts, demonstrating that bootstrap methods offer a practical trade-off between accuracy and computational cost, making them well-suited for dynamic intraday forecasting where rapid price fluctuations demand adaptive probabilistic modelling. 

In the BM, bootstrap methods have been explored to a lesser extent, but they show promise in improving forecast stability under extreme market conditions, as \cite{tahmasebifar2017point} demonstrates by employing bootstrap in a hybrid ensemble framework to generate probabilistic forecasts for both DAM and BM, improving prediction stability in highly volatile trading periods. Similarly, \cite{narajewski2022probabilistic} utilises bootstrap to generate probabilistic forecasts for German BM prices, showing that resampling techniques enhance empirical coverage under market uncertainty. Given the BM’s highly stochastic nature, the effectiveness of bootstrap methods in real-time imbalance price forecasting remains uncertain, as their computational feasibility may be challenged by the rapid adjustments required in balancing markets. However, bootstrap methods have proven effective in the DAM by enhancing probabilistic accuracy and forecast reliability, are emerging in the IDM for constructing adaptive PIs, and show potential in the BM for improving empirical coverage, though further research is needed to assess their ability to handle extreme real-time price volatility.

\subsubsection{Historical Simulation}
Historical simulation is a direct method of constructing probabilistic forecasts. It takes different names in the literature and is sometimes referred to as an empirical error distribution approach. This method explores both point forecasts and forecast errors. It is constructed as follows:
\begin{equation}
PI_{hist}^{1-\alpha} = \left[ \hat{P}_{d+1,h} + \gamma_{\alpha/2}, \hat{P}_{d+1,h} + \gamma_{1-\alpha/2} \right],
\end{equation}
where \(1-\alpha\) is the PI's nominal coverage level. The variables \(\gamma_{\alpha/2}\) and \(\gamma_{1-\alpha/2}\) represent the \((\alpha/2)\)-quantile and the \((1-\alpha/2)\)-quantile of \(\hat{\epsilon}_{d,h}\), respectively.

In the DAM, historical simulation has been used to generate probabilistic forecasts by leveraging past forecast errors to estimate empirical PIs. \cite{weron2008forecasting} applies this method to spot electricity prices, demonstrating that sample quantiles derived from one-step-ahead prediction errors effectively capture price uncertainty without relying on parametric assumptions. Building on this, \cite{nowotarski2015computing} employs historical simulation in combination with Autoregressive with Exogenous Features (ARX) and Semi-Nonparametric ARX (SNARX) residuals, showing that this approach is particularly effective in volatile market conditions where model-driven approaches may struggle. Further emphasising its practical advantages, \cite{uniejewski2019importance} compares historical simulation with bootstrapping and finds that it achieves similar forecasting performance while offering a simpler and more computationally efficient alternative. The robustness of historical simulation in capturing volatility and price spikes is also highlighted by \cite{janczura2024expectile}, who demonstrates its effectiveness in modelling extreme price movements without requiring parametric distributions. 

While historical simulation has been widely explored in the DAM, its application in the IDM has also shown promise, particularly in modelling intraday price volatility. \cite{marcjasz2020probabilistic} and \cite{maciejowska2024probabilistic} apply historical simulation for DAM and IDM price forecasting, with the former leveraging residual-based methods to enhance accuracy, while the latter shows it captures past volatility but underestimates uncertainty, particularly in the IDM with its erratic price dynamics. Further expanding on its multidimensional capabilities, \cite{maciejowska2024multiple} employs historical simulation in a multiple-output forecasting framework, constructing PIs by incorporating historical forecast errors into point forecasts. \cite{hirsch2024multivariate} develops a multivariate simulation-based forecasting framework for the German IDM, integrating copula-based models to capture cross-product price dependencies, demonstrating that historical simulation can improve PEPF accuracy in high-volatility trading environments where electricity contract dependencies influence price formation. 

Despite its success in DAM and IDM, historical simulation remains unexplored in the BM, where its empirical approach to constructing PIs could offer a viable alternative to parametric methods, though its ability to capture rapid price fluctuations and BM settlement dynamics in a highly volatile, real-time environment remains uncertain.

\subsection{Regression-Based Approaches}  
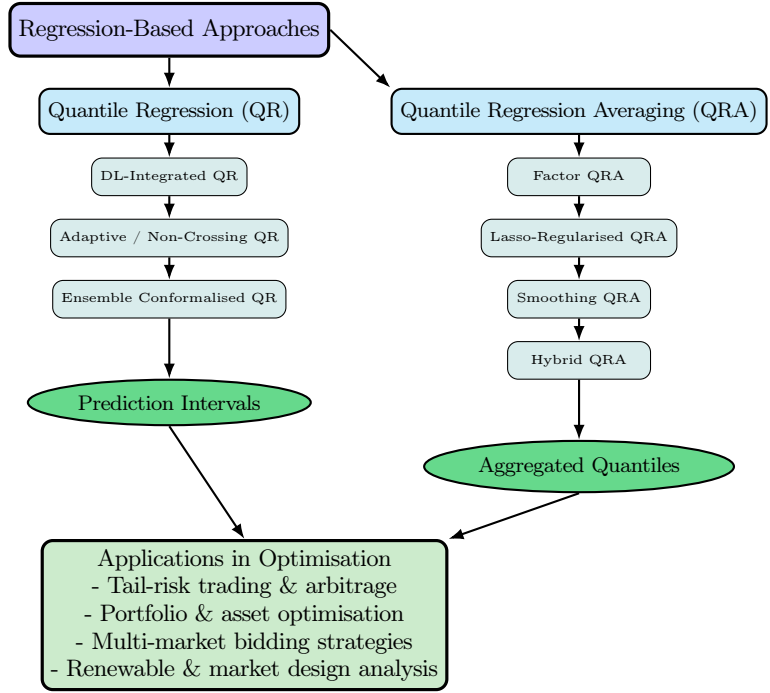
\begin{figure}[ht]
\centering
\begin{tikzpicture}[
  node distance=3mm and 0mm,
  family/.style={rectangle, rounded corners, draw=black, very thick, fill=blue!20,
                 minimum width=3.2cm, minimum height=0.7cm, font=\small, align=center},
  method/.style={rectangle, rounded corners, draw=black, thick, fill=cyan!20,
                 minimum width=2.4cm, minimum height=0.6cm, font=\footnotesize, align=center},
  submethod/.style={rectangle, rounded corners, draw=black, thin, fill=teal!15,
                    minimum width=1.9cm, minimum height=0.5cm, font=\tiny, align=center},
  output/.style={ellipse, draw=black, thick, fill=green!50!teal!60,
                 minimum width=2.2cm, minimum height=0.6cm, font=\footnotesize, align=center},
  final/.style={rectangle, rounded corners, draw=black, very thick, fill=green!60!black!20,
                minimum width=3.5cm, minimum height=0.8cm, font=\small, align=center},
  arrow/.style={-{Latex[length=2mm]}, thick}
]

\node[family] (regression) {Regression-Based Approaches};

\node[method, below=of regression, yshift=-1mm] (qr) {Quantile Regression (QR)};
\node[method, right=12mm of qr] (qra) {Quantile Regression Averaging (QRA)};

\node[submethod, below=of qr] (qr_dl) {DL-Integrated QR};
\node[submethod, below=of qr_dl] (qr_adapt) {Adaptive / Non-Crossing QR};
\node[submethod, below=of qr_adapt] (qr_cp) {Ensemble Conformalised QR};

\node[submethod, below=of qra] (f_qra) {Factor QRA};
\node[submethod, below=of f_qra] (l_qra) {Lasso-Regularised QRA};
\node[submethod, below=of l_qra] (s_qra) {Smoothing QRA};
\node[submethod, below=of s_qra] (hybrid_qra) {Hybrid QRA};

\node[output, below=of qr_cp, yshift=-5mm] (pi) {Prediction Intervals};
\node[output, below=of hybrid_qra, yshift=-5mm] (agg) {Aggregated Quantiles};

\node[final, below=15mm of pi, xshift=10mm] (decision) {Applications in Optimisation
\\- Tail-risk trading \& arbitrage
\\- Portfolio \& asset optimisation
\\- Multi-market bidding strategies
\\- Renewable \& market design analysis};

\draw[arrow] (regression.south) -- (qr.north);         
\draw[arrow] (regression.east) -- (qra.north west);    

\draw[arrow] (qr) -- (qr_dl);
\draw[arrow] (qr_dl) -- (qr_adapt);
\draw[arrow] (qr_adapt) -- (qr_cp);

\draw[arrow] (qra) -- (f_qra);
\draw[arrow] (f_qra) -- (l_qra);
\draw[arrow] (l_qra) -- (s_qra);
\draw[arrow] (s_qra) -- (hybrid_qra);

\draw[arrow] (qr_cp) -- (pi);
\draw[arrow] (hybrid_qra) -- (agg);

\draw[arrow] (pi.south) -- (decision.north);        
\draw[arrow] (agg.south) -- (decision.north east);  

\end{tikzpicture}
\caption{Regression-based approaches: Quantile Regression (QR) and Quantile Regression Averaging (QRA).}
\label{RegPlot}
\end{figure}

Quantile-based techniques, such as quantile regression and its extensions, focus on predicting the conditional quantiles of electricity prices, offering a straightforward way to estimate the uncertainty without relying on strong distributional assumptions.  An outline of the methods can be seen in Figure \ref{RegPlot}.

\subsubsection{Quantile Regression (QR)}\label{QuantileRegression}  
QR, introduced by \cite{koenker1978regression}, extends ordinary least squares by estimating conditional quantiles of the response variable, offering insights into uncertainty across different distribution points. Unlike traditional methods focused on the mean, QR models various parts of the distribution, making it particularly useful in capturing the impact of predictors on extreme price values, such as electricity price spikes during peak demand.
Given a linear model:
\begin{equation}
y_t = \mathbf{x}_t^\top \mathbf{\beta} + \epsilon_t,
\end{equation}
where \( y_t \) is the dependent variable, \( \mathbf{x}_t \) are predictors, and \( \epsilon_t \) is the error term, QR aims to estimate the conditional quantile function \( \hat{y}_t^\alpha \), which represents the value below which a certain percentage \( \alpha \) of \( y_t \) falls, given \( \mathbf{x}_t \). The QR estimator is obtained by solving:
\begin{equation}
\hat{\beta}_\alpha = \arg\min_{\mathbf{\beta}} \sum_{t=1}^N \rho_\alpha(y_t - \mathbf{x}_t^\top \mathbf{\beta}),
\end{equation}
where \( \rho_\alpha(u) \) is the check function defined as:
\begin{equation}
\rho_\alpha(u) = 
\begin{cases}
\alpha u, & \text{if } u \geq 0, \\
(\alpha - 1)u, & \text{if } u < 0.
\end{cases}
\end{equation}
QR's asymmetric penalty allows it to estimate different quantiles, providing flexibility in capturing price variability, especially during market extremes. The estimator is consistent and asymptotically normal, with its covariance matrix \( \mathbf{\Sigma}_\alpha \) given by:
\begin{equation}
\mathbf{\Sigma}_\alpha = \alpha(1 - \alpha) \left[ f_{y_t|\mathbf{x}_t}(\hat{q}_t^\alpha)^2 \cdot (X'X)^{-1} \right],
\end{equation}
where \( f_{y_t|\mathbf{x}_t}(\hat{q}_t^\alpha) \) is the conditional density, and \( (X'X)^{-1} \) represents the design matrix inverse.

In the DAM, QR has been used to enhance PEPF accuracy, particularly in modelling tail risks and non-linear price dependencies, with \cite{jonsson2014predictive} introducing a time-adaptive QR approach that improves tail prediction by fitting exponential distributions, outperforming models like GARCH in capturing extreme price variations. Similarly, \cite{do2015using} applies QR to both price and demand forecasting, demonstrating that the impact of market drivers varies across quantiles, reinforcing the complexity of electricity price dynamics. The effectiveness of QR in conditional price distribution modelling is further validated by \cite{nowotarski2015computing}, who finds that while QR provides informative PIs, its performance is generally enhanced when combined with model averaging techniques such as QRA. This is particularly evident in \cite{bello2016medium}, where QR is integrated with a fundamental market equilibrium model for medium-term DAM price forecasting in Spain, with the results showing that QR-based hybrid models improve risk estimation, particularly for extreme price variations influenced by high wind penetration. Beyond its role in probabilistic forecasting, QR has been widely used to analyse the sensitivity of electricity prices to market fundamentals, with \cite{hagfors2016modeling} and \cite{tzallas2022probabilistic} applying QR to UK electricity prices to quantify how factors such as gas and coal prices impact different quantiles, providing valuable insights for risk management. Similarly, \cite{maciejowska2020assessing} examines the effect of renewable energy integration on German DAM prices, demonstrating how wind and solar generation influence both price levels and variability across different quantiles. More recently, DL frameworks have been integrated with QR to refine probabilistic forecasting accuracy, with \cite{chaweewat2020electricity} employing QR alongside a ResNet model to generate asymmetrical PIs that improve performance during price spikes. In a related study, \cite{jensen2022ensemble} introduces Ensemble Conformalised QR, combining QR with CP to enhance forecast sharpness and reliability, particularly in volatile market conditions. Further methodological advancements include \cite{liu2023day}, who integrates QR with LSTNet and SHAP feature selection, refining probabilistic distributions using KDE, and \cite{zhang2023probabilistic}, who applies QR to capture non-linear relationships in price forecasting, improving interval sharpness without imposing restrictive distributional assumptions. QR has also been utilised in wholesale market design, with \cite{monjazeb2024wholesale} demonstrating its effectiveness in modelling price uncertainty in pay-as-bid markets, where regional and climatic variations influence clearing prices. The application of QR in hybrid DL frameworks has further improved forecast efficiency, as seen in \cite{xu2024novel}, where QR is combined with QR-LSTM and multi-objective tuna swarm optimisation to refine quantile forecasts under dynamic market conditions. In the UK market \cite{osone2025quantile} develops a generalisable QR-based framework, demonstrating stable PEPF and low pinball loss even under operational constraints that exclude previous-day price. \cite{chen2025outlier} develop an outlier-adaptive non-crossing QR model that integrates variational mode decomposition and conformal symmetry to enhance reliability, sharpness, and robustness of probabilistic forecasts under price spike conditions.

While QR has been extensively applied in DAM PEPF, its use in real-time electricity markets remains less developed, with the IDM seeing one and the BM seeing several notable applications, particularly in hybrid forecasting and trading strategy optimization. In the IDM \cite{yu2025orderfusion} propose a DL framework that processes raw order-book data and ensures quantile monotonicity via a hierarchical quantile head, outperforming traditional feature-based QR methods in intraday markets.

In the BM we see more, albeit concentrated applications. \cite{maciejowska2024multiple} enhances QR for PEPF in both DAM and BM by integrating it with multiple split CP, generating robust probabilistic forecasts that adapt to market uncertainties. The comparative study by \cite{o2024conformal} further evaluates QR’s performance relative to CP in the DAM, showing that while QR provides sharper forecasts, it struggles to maintain high empirical coverage, whereas CP ensures greater reliability. Extending this analysis, \cite{o2025conformal} explores QR’s role in hybrid ensembles for BM forecasting, finding that although QR effectively narrows interval width within CP-based approaches, it continues to face challenges in maintaining validity. A more applied perspective is offered by \cite{o2025optimising}, which optimises quantile forecasts for use in heuristic and mixed-integer linear programming trading strategies for electricity arbitrage in DAM and BM, demonstrating the economic viability of battery energy storage systems when using quantile-based forecasts for market participation.

Despite its strong presence in the DAM and emerging use in the BM, QR has yet to be extensively applied to the IDM, presenting a significant research gap. Given the IDM’s continuous trading structure, QR could offer valuable insights into price dynamics, particularly for modelling conditional price distributions and spreads relative to DAM prices. Its absence in IDM research highlights an important avenue for future exploration, especially in adaptive quantile-based methods for real-time forecasting.

\subsubsection{Quantile Regression Averaging (QRA)}
QRA, introduced by \cite{nowotarski2015computing}, utilises the combined outputs from \(M\) point forecasting models, \([f^1(\mathbf{x}_t), f^2(\mathbf{x}_t), \dots, f^M(\mathbf{x}_t)]\), applying QR to to estimate quantiles \(\hat{q}_t^\alpha\) and \(\hat{q}_t^{1-\alpha}\), defining the lower and upper bounds of the expected price distribution. By leveraging the entire set of model outputs, QRA is capable of capturing the impact of predictors across different regions of the price distribution. This method captures both central tendencies and tail behaviours, allowing it to adapt to market fluctuations and provide robust forecasts. The predictors \(\mathbf{x}_t\) can also include explanatory variables beyond point forecasts from these individual models. As such, QRA has gained prominence in PEPF literature due to its ability to generate both accurate and more reliable PIs.

In the DAM, QRA has been extensively applied to improve PEPF accuracy by aggregating point forecasts from diverse models. \cite{nowotarski2015computing} first introduces QRA for PEPF, demonstrating that it consistently outperforms individual models by effectively capturing conditional price distributions. Building on this, \cite{maciejowska2016probabilistic} develops Factor QRA (FQRA), which reduces model complexity through principal component analysis while maintaining high forecasting accuracy. The effectiveness of QRA in volatile market conditions is further validated by \cite{uniejewski2019importance}, who finds that QRA outperforms traditional probabilistic forecasting methods by combining QRs from multiple base models. A comparative study by \cite{serafin2019averaging} evaluates QRA against QR Machine (QRM), showing that while QRM is computationally efficient, QRA provides more robust and reliable PIs, particularly under high market uncertainty. Several refinements have been introduced to improve QRA’s forecasting performance. \cite{marcjasz2020probabilistic} highlights QRA’s ability to transform point forecasts into full probabilistic distributions, outperforming simpler methods such as historical simulation. The integration of DL models with QRA has further improved PI sharpness. \cite{chaweewat2020electricity} compares QRA against Residual NNs (ResNet-QR), noting that while QRA is effective, ResNet-QR produces narrower PIs with higher empirical coverage. Regularization techniques have also been incorporated to refine QRA models. \cite{uniejewski2021regularized} introduces Lasso-Regularized QRA (LQRA), demonstrating that penalizing irrelevant predictors reduces overfitting and improves forecast accuracy. \cite{maciejowska2022forecasting} demonstrates QRA’s robustness in capturing uncertainty and volatility in electricity markets, consistently providing accurate PIs. Similarly, \cite{uniejewski2023enhancing} and \cite{maciejowska2023probabilistic} explore LQRA and Factor QRA, showing that they enhance probabilistic forecasts by improving predictor selection and capturing latent market factors. More recently, \cite{uniejewski2023smoothing} introduces Smoothing QRA (SQRA) to address limitations in traditional QRA, improving both sharpness and reliability of probabilistic forecasts. In the German DAM \cite{lipiecki2025isotonic} benchmarks QRA with multiple variants, finding that while QRA provides competitive sharpness, it underperforms isotonic QRA and LQRA in both efficiency and validity metrics, with poor tail balance at high confidence levels.

While QRA has been widely adopted in the DAM, its presence in the IDM remains limited, with a few studies integrating it into hybrid forecasting approaches, such as \cite{kath2021conformal}, which combines QRA with CP to capture price variability while improving validity and empirical coverage. Similarly, \cite{maciejowska2024probabilistic} applies a hybrid Factor-QRA method for PEPF in both DAM and IDM, demonstrating improved forecast reliability in high-volatility trading environments, suggesting that while QRA can be adapted to continuous trading markets, further research is needed to assess its real-time forecasting accuracy. 

In the BM, QRA’s performance has been variable, with some studies highlighting its potential for improving probabilistic trading strategies, while others suggest that CP-based approaches provide superior performance in regards to validity. \cite{janczura2022dynamic} applies QRA to optimize short-term trading strategies across the DAM, IDM, and BM, finding that QRA-based probabilistic forecasts improve decision-making by balancing risk and profitability in multi-market trading. However, \cite{o2025conformal} compares QRA with CP for PEPF in DAM and BM, showing that while QRA produces sharper forecasts, CP-based methods consistently outperform it in terms of probabilistic coverage and financial gains.  Similarly, \cite{cornell2024probabilistic} applies QRA for probabilistic forecasting in the Australian National Electricity Market, showing that ensemble QRA models outperform individual models, especially when combined with post-processing techniques like spike filtration and autoregression. While QRA remains a strong baseline for BM forecasting, its effectiveness depends on market volatility and may require calibration to improve empirical coverage. 

Hybrid approaches have extended QRA’s capabilities in EPF, as \cite{marcjasz2023distributional} demonstrates that integrating QRA with ensemble learning techniques, including LASSO and DNNs, enhances probabilistic forecasts, though DDNNs outperform QRA when more flexible probabilistic distributions are needed. Similarly, \cite{feron2024probabilistic} explores the integration of QRA with Quantile Random Forests (QRF) and Conformalised Quantile Estimators, showing that hybrid models improve PIs, particularly in volatile markets. \cite{janczura2024expectile} compares QRA with Expectile Regression Averaging, finding both methods improve forecast accuracy, with QRA offering robust probabilistic forecasts.  More broadly, \cite{maciejowska2022forecasting} and \cite{lipiecki2024postprocessing} reaffirm QRA’s robustness in capturing uncertainty, particularly in ensemble forecasting frameworks. Further methodological enhancements have been explored in \cite{jiang2023electricity} and \cite{zakrzewski2024remodels}, who introduce nonconvex regularization techniques and dimensionality reduction methods to optimize QRA’s input selection and computational efficiency.  \cite{o2024conformal} compare QRA with CP for DAM PEPF, showing that while QRA provides reliable PI, CP techniques such as EnbPI and SPCI outperform QRA in terms of financial gains and coverage reliability in trading applications. Recent benchmarking studies, such as \cite{brusaferri2024line}, compare QRA with DDNNs, finding that while QRA remains a strong baseline, DL models with conformalised recalibration achieve superior coverage reliability and sharper PIs. The economic implications of QRA are further examined in \cite{le2025probablistic}, who apply QRA to DAM prices in the Czech market, demonstrating that smoothing the QRA objective function enhances forecast reliability while incorporating Lasso penalties improves input selection.

Having consistently outperformed individual models in DAM forecasting, QRA has evolved through factor-based variants, regularisation techniques, and integration with DL. However, its weaker empirical coverage in IDM and BM highlights the need for further refinement, particularly through hybrid and adaptive learning approaches, to enhance its real-time forecasting performance in more volatile, high-frequency markets.

\subsection{Conformal Prediction Methods}  
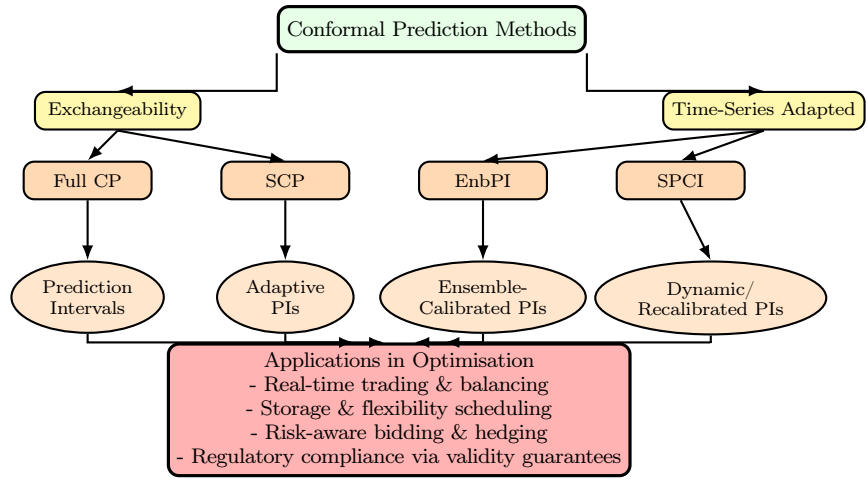
\begin{figure}[ht]
\centering
\begin{tikzpicture}[
  node distance=2mm and 3mm,
  family/.style={rectangle, rounded corners, draw=black, very thick, fill=green!10,
                 minimum width=2.5cm, minimum height=0.6cm, font=\footnotesize, align=center},
  layer/.style={rectangle, rounded corners, draw=black, thick, fill=yellow!40,
                 minimum width=2.2cm, minimum height=0.5cm, font=\scriptsize, align=center},
  method/.style={rectangle, rounded corners, draw=black, thick, fill=orange!30,
                 minimum width=1.7cm, minimum height=0.5cm, font=\scriptsize, align=center},
  output/.style={ellipse, draw=black, thick, fill=orange!20,
                 minimum width=1.8cm, minimum height=0.5cm, font=\scriptsize, align=center},
  final/.style={rectangle, rounded corners, draw=black, very thick, fill=red!30,
                minimum width=2.5cm, minimum height=0.7cm, font=\footnotesize, align=center},
  arrow/.style={-{Latex[length=2mm]}, thick}
]

\node[family] (cp) {Conformal Prediction Methods};

\node[layer, below left=of cp, xshift=-7mm, yshift=-3mm] (exch) {Exchangeability};
\node[layer, below right=of cp, xshift=7mm, yshift=-3mm] (ts) {Time-Series Adapted};

\node[method, below=of exch, xshift=-4mm, yshift=-2mm] (fullcp) {Full CP};
\node[method, right=of fullcp, xshift=6mm] (scp) {SCP};
\node[method, right=of scp, xshift=6mm] (enbpi) {EnbPI};
\node[method, right=of enbpi, xshift=6mm] (spci) {SPCI};

\node[output, below=of fullcp, yshift=-6mm] (pi_full) {Prediction\\Intervals};
\node[output, below=of scp, yshift=-6mm] (pi_scp) {Adaptive\\PIs};
\node[output, below=of enbpi, yshift=-6mm] (pi_enbpi) {Ensemble-\\Calibrated PIs};
\node[output, below=of spci, yshift=-6mm, xshift=4mm] (pi_spci) {Dynamic/\\Recalibrated PIs};

\node[final] (decision) at ($(pi_full.south)!0.5!(pi_spci.south) + (0,-10mm)$) {Applications in Optimisation
\\- Real-time trading \& balancing
\\- Storage \& flexibility scheduling
\\- Risk-aware bidding \& hedging
\\- Regulatory compliance via validity guarantees};

\draw[arrow] (cp.south west) |- (exch.north);
\draw[arrow] (cp.south east) |- (ts.north);

\draw[arrow] (exch.south) -- (fullcp.north);
\draw[arrow] (exch.south) -- (scp.north);
\draw[arrow] (ts.south) -- (enbpi.north);
\draw[arrow] (ts.south) -- (spci.north);

\draw[arrow] (fullcp.south) -- (pi_full.north);
\draw[arrow] (scp.south) -- (pi_scp.north);
\draw[arrow] (enbpi.south) -- (pi_enbpi.north);
\draw[arrow] (spci.south) -- (pi_spci.north);

\draw[arrow] (pi_full.south) -- ++(0,-1mm) |- ([xshift=-6mm]decision.north);
\draw[arrow] (pi_scp.south) -- ++(0,-1mm) |- ([xshift=-2mm]decision.north);
\draw[arrow] (pi_enbpi.south) -- ++(0,-1mm) |- ([xshift=2mm]decision.north);
\draw[arrow] (pi_spci.south) -- ++(0,-1mm) |- ([xshift=6mm]decision.north);

\end{tikzpicture}
\caption{Conformal Prediction methods: Full Conformal Prediction (Full CP), Split Conformal Prediction (SCP), Ensemble Batch Predictive Interval (EnbPI), and Sequential Predictive Conformal Inference (SPCI).}
\label{ConformPlot}

\end{figure}

CP offers a distribution-free framework for uncertainty quantification, providing a compelling alternative to QR and QRA. Originally introduced by \cite{gammerman1998learning} and later extended by \cite{vovk2005algorithmic} and \cite{shafer2008tutorial}, CP constructs PIs with valid coverage guarantees, regardless of the underlying data distribution.  An outline of the methods can be seen in Figure \ref{ConformPlot}.
At the core of CP is the concept of a \textit{nonconformity measure}, which quantifies how unusual or atypical a new data point \((\mathbf{x}_t, y_t)\) appears in relation to a dataset \(\mathcal{Z}\). For regression tasks, a common nonconformity score is the absolute prediction error \(A(\mathbf{x}_t, y_t) = |y_t - f(\mathbf{x}_t)|,\) where \(f(\mathbf{x}_t)\) is the predicted value. The nonconformity score measures how far the predicted value deviates from the true value. Other measures may be used depending on the specific model or application.
After computing nonconformity scores for all examples in \(\mathcal{Z}\), CP constructs a PI for a new observation \(\mathbf{x}_{\text{new}}\). For a candidate label \(y_{\text{new}}\), the corresponding p-value is calculated as:
\begin{equation}
p(y_{\text{new}}) = \frac{|\{(\mathbf{x}_t, y_t) \in \mathcal{Z} : A(\mathbf{x}_t, y_t) \geq A(\mathbf{x}_{\text{new}}, y_{\text{new}})\}| + 1}{N + 1}.
\end{equation}
This p-value reflects the proportion of past nonconformity scores that are greater or equal to the score for \(y_{\text{new}}\). The PI for \(\mathbf{x}_{\text{new}}\) is constructed by including all candidate labels \(y_{\text{new}}\) whose p-values exceed a predefined significance level \(\alpha\):
\begin{equation}
\Gamma_{1-2\alpha}(\mathbf{x}_{\text{new}}) = \{y_{\text{new}} : p(y_{\text{new}}) > \alpha\}.
\end{equation}
CP provides a rigorous validity guarantee, ensuring that with probability at least \(1 - 2\alpha\), the true response \(y_{\text{new}}\) will lie within the PI, \(\Gamma_{1-2\alpha}(\mathbf{x}_{t})\). This validity property holds regardless of the underlying data distribution, making CP effective when the distribution is unknown or difficult to model.
The PI width is determined by the quantile \(\lambda_t\) of the nonconformity scores where \(
\lambda_{t} = \text{Quantile}_{\alpha} \left( A_1, A_2, \dots, A_N \right), \) for the \(\alpha\) quantile level, with \(\lambda_t\) corresponding to the \((1-\alpha)(N+1)\)-th smallest nonconformity score in the sorted list. 

In the DAM, CP has been leveraged to refine probabilistic forecasts by ensuring reliable uncertainty quantification across different price conditions. \cite{tibshirani2019conformal} extends the CP framework by introducing Weighted CP (WCP), which accounts for covariate shifts and ensures that PIs remain valid even under changing market distributions. Applying CP to electricity price forecasting, \cite{bellomi2022prediction} demonstrates that seasonally adjusted price data significantly enhances the sharpness and reliability of PIs in the Nord Pool DAM, particularly in markets characterized by lower variance and fewer extreme price events. The flexibility of CP in transforming point forecasts into probabilistic predictions is further emphasized by \cite{dewolf2023valid}, who highlights its model-agnostic nature and ability to adapt to a wide range of forecasting methodologies. The integration of CP with other probabilistic forecasting techniques has further improved its forecasting performance. \cite{jensen2022ensemble} combines CP with QR in Ensemble Conformalized QR (EnCQR), demonstrating that this hybrid approach enhances the sharpness of PIs while maintaining empirical validity. Similarly, \cite{cordier2023flexible} develops a general CP-based forecasting framework, incorporating methods such as split-CP and Conformalized QR (CQR) to dynamically adjust PIs for time series forecasting. DL-based approaches have also benefited from CP-based recalibration. \cite{wang2023conformal} integrates CP with generative transformer models, producing reliable and narrow PIs for wind power forecasting—an approach that could be extended to electricity price forecasting in renewable-dominated markets. The postprocessing capabilities of CP have also been explored in \cite{lipiecki2024postprocessing}, who apply CP Averaging (CPA) to DAM prices, demonstrating its ability to adaptively refine PIs and improve forecast reliability. Furthermore, \cite{brusaferri2024line} enhances CP by incorporating neural network ensembles and CQR, showing that these refinements significantly improve the robustness of PIs in volatile electricity markets. In the German market \cite{lipiecki2025isotonic} benchmarks CP against several postprocessing methods, finding that while CP offers strong coverage, it underperforms isotonic QRA and LQRA in sharpness and tail balance, especially at lower confidence levels.

While CP has been widely applied in the DAM, its role in real-time electricity markets is still emerging. In the IDM, \cite{maciejowska2024probabilistic} applies CP to the German EPEX DAM and IDM, demonstrating that it consistently produces reliable PIs that maintain empirical coverage close to nominal levels. Their results show that CP-based forecasts frequently pass the Christoffersen test for conditional coverage, indicating that CP effectively captures price uncertainty in high-frequency intraday markets, though further work is needed to improve performance under extreme price swings. 

In the BM, CP remains underexplored, but \cite{alghumayjan2024conformal} show it can effectively quantify uncertainty and improve risk-adjusted returns for energy storage arbitrage in the New York real-time market, highlighting its potential for enhancing decision-making under extreme price volatility, though further work is needed to assess its integration with real-time signals and computational demands.

Overall, CP has proven to be a robust probabilistic forecasting tool, offering guaranteed empirical coverage and adaptability across model types. While its effectiveness in the DAM is well established, its application in the IDM and BM remains limited, highlighting important opportunities for future research in real-time and high-frequency market settings.

\subsubsection{Split Conformal Prediction (SCP)}
SCP is an inductive variant of CP that improves computational efficiency by dividing the dataset $\mathcal{Z}$ into a training set of size $n_{\text{train}}$ and a calibration set of size \(n_{\text{cal}}\), where \(N = n_{\text{train}} + n_{\text{cal}}\). By restricting nonconformity score calculations to the calibration set, SCP reduces the need to recalculate scores across the entire dataset, significantly cutting computational time and making it ideal for larger datasets. First, a model is trained on the training set to estimate the conditional expectation $f(\mathbf{x}_t)$. Nonconformity scores for the calibration set are then computed using the defined nonconformity score \(A(\mathbf{x}_t, y_t) = |y_t - f(\mathbf{x}_t)|\). 
To construct PIs for a new input \(\mathbf{x}_{\text{new}}\), SCP selects \(\lambda_t\) corresponding to the \((1-\alpha)(N+1)\)-th smallest nonconformity score computed from the calibration set, such that the PI for a new input $\mathbf{x}_{t}$ is given by:
\begin{equation}
\Gamma_{1-2\alpha}(\mathbf{x}_{t}) = \left[ f(\mathbf{x}_{t}) - \lambda_{t}, f(\mathbf{x}_{t}) + \lambda_{t} \right]
\end{equation}
providing a probabilistic guarantee that $y_{t}$ lies within this range with a confidence level of $1 - 2\alpha$, assuming data exchangeability. In cases where the exchangeability assumption is violated, such as in time-series data like electricity prices, adaptations EnbPI and SPCI can better handle temporal dependencies.

In the DAM, SCP has been widely applied to enhance PEPF by constructing valid and adaptive PIs. Early studies, such as \cite{kath2021conformal}, apply SCP to DAM and IDM markets, demonstrating that SCP produces valid and sharp PIs, particularly in volatile trading conditions. However, the effectiveness of SCP depends on data partitioning strategies, with methods like One-Split SCP providing dynamically adjusted PIs for non-stationary electricity prices \cite{dutot2024adaptive}. Recent advancements have focused on hybrid SCP approaches—for instance, \cite{hu2022conformalized} combines SCP with QR in a Temporal Convolutional Network framework for wind power forecasting, ensuring that PIs meet coverage requirements over time. Similarly, \cite{feron2024probabilistic} introduces one split SCQR, a method integrating SCP with CQR, improving SCP’s adaptability in time-series forecasting. Beyond QR-based integrations, SCP has also been extended through ensemble and multidimensional approaches, with \cite{maciejowska2024multiple} developing Multiple-Split SCP to construct multidimensional PIs that enhance uncertainty quantification by leveraging multiple calibration sets. SCP has also been applied in distribution-based methods, with \cite{jonkers2024novel} introducing Split Conformal Distribution Regression Forests to refine SCP’s efficiency in probabilistic forecasting. While SCP has been successfully implemented in PEPF for DAM, its performance relative to other CP methods remains mixed, as \cite{o2024conformal} demonstrates in the Irish DAM, showing that although SCP ensures valid coverage, it is outperformed by EnbPI and SPCI in sharpness and financial trading performance. 

Extending this analysis, \cite{o2025conformal} evaluates SCP in both DAM and BM, finding that while SCP maintains strong coverage, it struggles in the highly volatile BM, where EnbPI and SPCI provide narrower PIs and hold superior financial performance due to their ability to overcome SCP’s reliance on exchangeability assumptions.

Despite SCP’s success in the DAM, and to a lesser extent in the BM, it remains unexplored in the IDM, where continuous trading structures pose additional challenges. Its limited use in PEPF highlights the need for further refinements to improve its applicability in real-time and intraday settings, particularly by addressing violations of the exchangeability assumption.

\subsubsection{Ensemble Batch Predictive Interval (EnbPI)}\label{EnbPI}
EnbPI, introduced by \cite{xu2021conformal}, builds on the CP framework, providing PIs with finite-sample coverage guarantees. Unlike traditional CP methods, EnbPI leverages ensemble learning and bootstrap sampling to generate reliable PIs without the need for separate calibration sets, making it particularly effective in time-series forecasting applications with complex dependencies.
Utilising the dataset \(\mathcal{Z}\), EnbPI generates \(B\) bootstrap samples \(\mathcal{Z}^{(b)}\), where each sample trains a predictive model \(f^{(b)}(\mathbf{x}_{t})\), resulting in multiple separate models \( \{f^{(1)}(\mathbf{x}_{t}), f^{(2)}(\mathbf{x}_{t}), \dots, f^{(B)}(\mathbf{x}_{t})\}\), with residuals calculated as \(e_t^{(b)} = |y_t - f^{(b)}(\mathbf{x}_t)|\). Thus the interval \(\lambda^{(b)}_{t}\) from these residuals, \(\lambda^{(b)}_{t} = \text{Quantile}_{\alpha}(e_1^{(b)}, e_2^{(b)}, \dots, e_N^{(b)}),\) together with the average of the ensemble predictions \( f^{(b)}(\mathbf{x}_{t}) \), forms the PI:
\begin{equation}
\Gamma_{1-2\alpha}(\mathbf{x}_{t}) = \left[ \frac{1}{B} \sum_{b=1}^B f^{(b)}(\mathbf{x}_{t}) - \frac{1}{B} \sum_{b=1}^B \lambda^{(b)}_{t}, \frac{1}{B} \sum_{b=1}^B f^{(b)}(\mathbf{x}_{t}) + \frac{1}{B} \sum_{b=1}^B \lambda^{(b)}_{t} \right],
\end{equation}
This construction allows EnbPI to offer valid coverage without retraining models, and by leveraging leave-one-out residuals, ensuring minimal computational.

EnbPI has been widely adopted in PEPF due to its ability to generate reliable PIs without making strong distributional assumptions. \cite{jensen2022ensemble} applies EnbPI as a CP method that leverages ensemble learning to improve forecast validity while dynamically adapting to market conditions. Expanding on this, \cite{zaffran2022adaptive} applies EnbPI to PEPF, demonstrating that its block bootstrap resampling enhances coverage stability by recalibrating PIs based on recent data, effectively capturing seasonality and trends. The methodological flexibility of EnbPI is further highlighted by \cite{cordier2023flexible}, who incorporates EnbPI into the MAPIE library, emphasising its ability to maintain valid coverage in non-exchangeable time-series data. In market-specific applications, \cite{de2024electricity} evaluates EnbPI in the Mexican wholesale electricity market, showing that bootstrap-based ensemble estimators effectively capture price volatility and achieve the target 95\% coverage level. Within European electricity markets, \cite{o2024conformal} applies EnbPI to the Irish DAM, demonstrating that it outperforms traditional QR and SCP methods by producing narrower PIs while maintaining consistent coverage. 

Extending this analysis, \cite{o2025conformal} evaluates EnbPI in the Irish BM, where its maintains valid PIs while delivers stronger financial performance in ensemble-based approaches compared to standalone QR or SPCI models. 

Despite its strengths, EnbPI’s reliance on bootstrap resampling can introduce notable computational overhead. Its effectiveness in high-frequency trading environments such as the IDM and BM also remains underexplored, highlighting important directions for future research.

\subsubsection{Sequential Predictive Conformal Inference (SPCI)}\label{SPCI}
SPCI, introduced by \cite{xu2023sequential}, improves on EnbPI by dynamically adjusting PIs to accommodate new observations in time-series data. SPCI uses a sequential updating mechanism that adapts to evolving data distributions, ensuring reliable uncertainty quantification even in non-stationary environments like EPF.
Unlike static CP approaches, which recalibrate PIs using a fixed set of past data, SPCI updates non-conformity scores in real-time, leveraging the information from new residuals. For each new observation \((\mathbf{x}_{t}, y_{t})\), SPCI adjusts the PIs based on the residuals \( \epsilon_t\). This feedback loop enables SPCI to refine its predictions continuously, ensuring that the intervals remain accurate and informative.

SPCI provides theoretical guarantees of asymptotic conditional coverage, ensuring that the probability of the true value \( y_{t+1} \) lying within the constructed interval converges to the desired confidence level \( 1 - \alpha \) over time, such that:
\begin{equation}
\lim_{t \to \infty} \mathbb{P}\left( y_{t+1} \in \Gamma_{1 - \alpha}(\mathbf{x}_{t+1}) \mid \mathbf{X}_t \right) \geq 1 - 2\alpha,
\end{equation}
where \( \mathbf{X}_t \) represents all information available up to time \( t \). 

The construction of the PI for a new input \( \mathbf{x}_{t} \) is given by:
\begin{equation}
\Gamma_{1 - 2\alpha}(\mathbf{x}_{t}) = \left[ f(\mathbf{x}_{t}) - \lambda_t, f(\mathbf{x}_{t}) + \lambda_t \right],
\end{equation}
where \( \lambda_t \) is the quantile of the residuals up to time \( t \).  By using a sliding window of recent residuals and modelling their quantiles through QR Forests, SPCI efficiently captures changes in data, advancing time-series forecasting with theoretical coverage guarantees and producing narrower, more responsive PIs for reliable uncertainty quantification.

SPCI has emerged as a powerful extension of CP, offering dynamically recalibrated PIs that adapt to evolving market conditions. \cite{xu2024conformal} introduces MultiDimSPCI, extending SPCI to multivariate time series forecasting by constructing ellipsoidal prediction regions, ensuring valid coverage in high-dimensional, non-exchangeable settings. Further advancing its adaptability, \cite{angelopoulos2024conformal} enhances SPCI with Conformal PID Control, enabling the method to adjust PIs in response to distribution shifts and long-term trends, thereby improving robustness in dynamic environments. Building on these refinements, \cite{lee2024transformer} presents SPCI-T, integrating the Transformer decoder to capture long-range dependencies, significantly improving both interval accuracy and coverage, particularly in non-stationary and multi-step forecasting scenarios. In PEPF, \cite{o2024conformal} applies SPCI to the Irish DAM, demonstrating that it outperforms QR and SCP by offering superior empirical coverage and sharper PIs while effectively quantifying market uncertainty. 

In the BM we see one notable application from \cite{o2025conformal} who benchmarks SPCI in the Irish BM, showing that it achieves narrower PIs with valid coverage more consistently than SCP and EnbPI. Its adaptive recalibration mechanism enhances reliability in non-stationary market conditions, leading to improved financial performance in trading strategies. Notably, when integrated into an ensemble with QR and EnbPI, SPCI provides the strongest financial returns, balancing interval efficiency with robust uncertainty quantification. 

Despite these advantages, much like CP, SCP, and EnbPI, further research is needed to assess its applicability in IDM settings, where its ability to adapt to continuous trading environments again remains unexplored.

\section{Discussion}\label{Discussion}
The following section delves into key trends in the PEPF literature by systematically examining methodological preferences, market coverage, and empirical performance. Drawing on the studies discussed in \ref{Uncertainty_Review} and outlined in Tables \ref{TableUELR} \& \ref{Table_non_UELR}, we identify shifts in forecasting paradigms, including the growing adoption of CP, the continued use of quantile-based approaches, and the gradual decline of more traditional techniques such as Bayesian inference and Monte Carlo simulation. Particular attention is paid to under-represented markets, namely the IDM and BM, and to emerging hybrid and ensemble frameworks that offer new avenues for handling forecast uncertainty in real time. This discussion also incorporates cross-market comparisons, validity-efficiency trade-offs, and implications for practical deployment, culminating in a set of targeted research directions for advancing the field.

\subsection{Overview}
Table \ref{TableUELR} presents a structured overview of the literature on PEPF, categorising studies by forecasting methodology and market focus. It highlights the widespread use of key approaches such as QR, QRA, CP, and Bayesian methods, and their deployment across electricity markets including the DAM, IDM, and RT/BM. The timeline of included studies reveals notable shifts in research priorities, particularly the growing adoption of CP in recent years. In Table \ref{TableUELR} we see clear trends in methodological preferences and market coverage over time. Earlier studies, pre-2020, frequently relied on Bayesian inference and MC simulation, which were well-suited for capturing uncertainty distributions and scenario analysis. However, these methods have seen a marked decline, supplanted by more flexible, data-driven alternatives such as ensemble models, DL, and frequentist techniques like CP. Bootstrap and DB methods also appear in the literature but are largely confined to specialised use cases rather than serving as dominant forecasting tools. HS remains the least common, likely due to its reliance on static empirical distributions and poor adaptability to real-time changes. Since GEFCom2014, quantile-based models, particularly QR and QRA, have emerged as leading approaches, with QRA now widely favoured for its ability to aggregate diverse model outputs and deliver more robust probabilistic forecasts.
\begin{table}[H]
\centering
\scriptsize 
\begin{tabular}{p{3.2cm} p{0.3cm} p{0.4cm} p{0.3cm} p{0.3cm} p{0.3cm} p{0.4cm} p{0.3cm} p{0.3cm} p{0.4cm} p{0.4cm}  p{0.4cm}}
\toprule
\textbf{Reference} & \textbf{QR} & \textbf{QRA} & \textbf{CP} & \textbf{BY} & \textbf{BS} &  \textbf{DB} & \textbf{MC}  & \textbf{HS} & \textbf{DAM} & \textbf{IDM}  & \textbf{RT}  \\
\midrule
Panagiotelis et al. 2008 \cite{panagiotelis2008bayesian} &    &    &    & \checkmark   &    &    &  \checkmark  &    &      &  \checkmark    & \\
Weron et al. 2008 \cite{weron2008forecasting}            &    &    &    &    &    &  \checkmark  &    & \checkmark   &  \checkmark   &       & \\

Janczura et al. 2010 \cite{janczura2010empirical}        &    &    &    &    &    &   \checkmark  &    &    &  \checkmark    &      &  \\
Vahidinasab et al. 2010 \cite{vahidinasab2010bayesian}   &    &    &    & \checkmark   &    &    & \checkmark   &    &   \checkmark   &      &  \\
Wu et al. 2010 \cite{wu2010hybrid}                       &    &    &    &    &    &    &  \checkmark   &    &   \checkmark  &       & \\
Brolin et al. 2010 \cite{brolin2010modeling}    &    &    &    &    &    &    &   \checkmark  &    &   &   & \checkmark \\

Serinaldi et al. 2011 \cite{serinaldi2011distributional} &    &    &    &    &    &  \checkmark   &    &    &    \checkmark  &      &  \\
Alonso et al. 2011 \cite{alonso2011seasonal}             &    &    &    &    & \checkmark   &    &  \checkmark  &    &   \checkmark  &      &  \\

Chen et al. 2012 \cite{chen2012electricity}             &    &    &    &    &  \checkmark  &    &    &    &    &   & \checkmark \\

Wan et al. 2013 \cite{wan2013hybrid}                     &    &    &    &    &  \checkmark  &    &    &    &    &   & \checkmark \\
Khosravi et al. 2013 \cite{khosravi2013quantifying}      &    &    &    &    &  \checkmark  &    &    &    &   &   & \checkmark \\

Jonsson et al. 2014 \cite{jonsson2014predictive}         &  \checkmark  &    &    &    &    &   \checkmark &    &    &   \checkmark  &      &  \\

Nowotarski et al. 2015 \cite{nowotarski2015computing}    &  \checkmark  &  \checkmark  &  &    &    &  &  &  \checkmark  &  \checkmark   &    &   \\
Do et al. 2015 \cite{do2015using}    &  \checkmark   &    &    &    &    &    &    &    &  \checkmark   &    & \\

Maciejowska et al. 2016 \cite{maciejowska2016probabilistic} &    &  \checkmark  &    &    &    &    &    &    &  \checkmark    &     &   \\
Hagfors et al. 2016 \cite{hagfors2016modeling}           &  \checkmark  &    &    &    &    &    &    &    & \checkmark    &      &  \\
Dudek et al. 2016 \cite{dudek2016multilayer}             &   \checkmark  &    &    &    &    &    &    &    &  \checkmark    &     &   \\
Bello et al. 2016 \cite{bello2016medium}                 &  \checkmark  &    &    &    &    &    & \checkmark   &    &  \checkmark   &     &   \\
Bello et al. 2016 \cite{bello2016probabilistic}          &    &    &    &    &    &    &   \checkmark  &    &  \checkmark  &     &   \\

Tahmasebifar et al. 2017 \cite{tahmasebifar2017point}    &    &    &    &    & \checkmark   &    &    &    &  \checkmark  &     &  \checkmark  \\

Tat et al. 2018 \cite{tat2018electricity}                &    &    &    &    &    &    &  \checkmark  &    &   \checkmark  &     &   \\
Nowotarski et al. 2018 \cite{nowotarski2018recent}       & \checkmark   & \checkmark   &    &  \checkmark  &  \checkmark  &    &  \checkmark  &  \checkmark  & \checkmark  &     &   \\
Ziel et al. 2018 \cite{ziel2018probabilistic}            & \checkmark   &    &    &    &    & \checkmark   &    &    &  \checkmark  &      &  \\
Dudek et al. 2018 \cite{dudek2018probabilistic}          &    &    &    &    &    &  \checkmark  &    &    &  \checkmark  &    &    \\
Monteiro et al. 2018 \cite{monteiro2018new}              &    &    &    &    &    &  \checkmark  &    &    &  \checkmark  &     &   \\
Osorio et al. 2018 \cite{osorio2018hybrid}               &    &    &    &    &    &    &   \checkmark  &    & \checkmark   &     &   \\

Dumas et al. 2019 \cite{dumas2019probabilistic}          &    &    &    &    &    &  \checkmark    &    &    &    &     &  \checkmark \\
Kostrzewski et al. 2019 \cite{kostrzewski2019probabilistic} &    &    &    &  \checkmark  &    &    &    &    &   \checkmark   &    &   \\
Brusaferri et al. 2019 \cite{brusaferri2019bayesian}     &    &    &    &  \checkmark  &    &    &    &    &   \checkmark   &    &   \\
Cheng et al. 2019 \cite{cheng2019hybrid}                 &    &    &    & \checkmark   &    &    &    &    &  \checkmark   &     &   \\
Uniejewski et al. 2019 \cite{uniejewski2019importance}   &    &   \checkmark  &    &    & \checkmark    &    &    & \checkmark    &   \checkmark   &     &   \\
Serafin et al. 2019 \cite{serafin2019averaging}          &    & \checkmark   &    &    &    &  \checkmark  &    &    &   \checkmark &     &   \\
Muller et al. 2019 \cite{muller2019bayesian}             &    &    &    &  \checkmark  &    &    &    &    &  \checkmark  &    &    \\
Ghayekhloo et al. 2019 \cite{ghayekhloo2019combination}  &    &    &    &  \checkmark  &    &    &    &    &  \checkmark  &    &    \\

Marcjasz et al. 2020 \cite{marcjasz2020probabilistic}    &    &  \checkmark  &    &    &    &    &    &  \checkmark  &  \checkmark    &     &   \\
Chaweewat et al. 2020 \cite{chaweewat2020electricity}    &  \checkmark  & \checkmark   &    &    &    &    &    &    &  \checkmark  &    &    \\
Brusaferri et al. 2020 \cite{brusaferri2020probabilistic}  &    &    &    & \checkmark   &    &    &    &    &     &     \\
Maciejowska et al. 2020 \cite{maciejowska2020assessing}    &  \checkmark  &    &    &    &    &    &    &    &  \checkmark  &   &   \\

Uniejewski et al. 2021 \cite{uniejewski2021regularized} &    &  \checkmark  &    &    &    &    &    &    &      &       &   \\
Kath et al. 2021 \cite{kath2021conformal}               &    &  \checkmark   &  \checkmark  &   &    &    &    &     &  \checkmark    &  \checkmark      &   \\
Bhatia et al. 2021 \cite{bhatia2021ensemble}            &    &    &    &  \checkmark   & \checkmark    &    &    &    &     &       &   \\
\end{tabular}
\caption{Literature on PEPF. Acronyms- Bayesian (BY), Bootstrap (BS), Distributional Based (DB), Monte Carlo (MC), Historical Simulation (HS),  Real-Time (RT). The \checkmark is the approach presented herein.}
\label{TableUELR}
\end{table}
CP has gained significant traction since 2021, particularly for its ability to provide empirical coverage guarantees and adapt to non-stationary, time-series settings. Its role as a recalibration layer has been especially prominent in post-2020 hybrid frameworks that integrate CP with QRA, QR, or ensemble methods to balance sharpness and validity. Among these, QRA remains the most dominant technique, consistently outperforming standalone QR in recent benchmarks. Overall, the literature reflects a shift from rigid probabilistic assumptions to flexible, hybrid approaches capable of capturing both model and distributional uncertainty in complex electricity markets, as well as meeting validity requirements.
\begin{table}[H]
\centering
\scriptsize 
\begin{tabular}{p{3.2cm} p{0.3cm} p{0.4cm} p{0.3cm} p{0.3cm} p{0.3cm} p{0.4cm} p{0.3cm} p{0.3cm} p{0.4cm} p{0.4cm}  p{0.4cm}}
\toprule
\textbf{Reference} & \textbf{QR} & \textbf{QRA} & \textbf{CP} & \textbf{BY} & \textbf{BS} &  \textbf{DB} & \textbf{MC}  & \textbf{HS} & \textbf{DAM} & \textbf{IDM}  & \textbf{RT}  \\
\midrule
Narajewski et al. 2022 \cite{narajewski2022probabilistic} &    &    &    &    &  \checkmark  &  \checkmark  &    &    &     &   \checkmark   &    \\
Narajewski et al. 2022 \cite{narajewski2022optimal}     &    &    &    &    &  \checkmark  &    &    &  \checkmark  &    &    &   \\
Maciejowska et al. 2022 \cite{maciejowska2022forecasting}  &    & \checkmark   &    &    &    &    &  \checkmark  & \checkmark    &   &    &     \\
Gabrielli et al. 2022 \cite{gabrielli2022data}          &    &    &    &    &    &    &   \checkmark  &    &     &    &     \\
Yildirim et al. 2022 \cite{yildirim2022supply}          &    &    &    &    &    &    &  \checkmark  &    &    &   &   \\
Lima et al. 2022 \cite{lima2022bayesian}                &    &    &    &  \checkmark  &    &    &    &    &    &       &  \checkmark   \\
Tzallas et al. 2022 \cite{tzallas2022probabilistic}     &  \checkmark  &    &    &    &    &    &    &    &   \checkmark &       &    \\
Janczura et al. 2022 \cite{janczura2022dynamic}         &    &  \checkmark  &    &    &    &    &    &    &  \checkmark  & \checkmark      & \checkmark   \\
Bellomi et al. 2022 \cite{bellomi2022prediction}        &    &    &  \checkmark    &    &    &    &    &    &    \checkmark  &       &    \\

Yuanchen et al. 2023 \cite{yuanchen2023electricity}      &    &    &    &  \checkmark   &    &    &    &    &    &    &  \checkmark  \\
Klein et al. 2023 \cite{klein2023deep}                   &    &    &    &  \checkmark  &    &  \checkmark   &  \checkmark  &    &    & \checkmark    &   \\
Uniejewski et al. 2023 \cite{uniejewski2023smoothing}    &    &  \checkmark  &    &    &    &    &    &    &    \checkmark &     &   \\
Uniejewski et al. 2023 \cite{uniejewski2023enhancing}      &    &  \checkmark  &    &    &    &    &    &    &  \checkmark  &     &  \\
Marcjasz et al. 2023 \cite{marcjasz2023distributional}   &    &  \checkmark  &    &    &    &  \checkmark  &    &    &  \checkmark  &    &   \\
Liu et al. 2023 \cite{liu2023day}                        &  \checkmark  &    &    &    &    &    &    &    &  \checkmark   &   &   \\
Zhang et al. 2023 \cite{zhang2023probabilistic}          &  \checkmark  &    &    &    &   \checkmark &    &    &   & \checkmark   &     &    \\
Maciejowska et al. 2023 \cite{maciejowska2023probabilistic}  &    &  \checkmark  &    &    &    &    &    &    & \checkmark    &    &    \\
Cramer et al. 2023 \cite{cramer2023multivariate}             &    &    &    &    &    &  \checkmark  &    &    &  \checkmark  &  \checkmark &   \\
Jiang et al. 2023 \cite{jiang2023electricity}             &    &  \checkmark  &    &    &    &    &    &    &  \checkmark  &   \checkmark  &   \\

Nickelsen et al. 2024 \cite{nickelsen2024bayesian}      &    &    &    &   \checkmark &    &    &    &    &    &  \checkmark   &   \\
Monjazeb et al. 2024 \cite{monjazeb2024wholesale}       &  \checkmark  &    &    &    &    &    &    &    &  \checkmark   &    &   \\
De et al. 2024 \cite{de2024electricity}                 &    &    &  \checkmark  &    &    &    &    &    &  \checkmark   &     &    \\
Xu et al. 2024 \cite{xu2024novel}                         &  \checkmark  &    &    &    &    &  \checkmark  &    &   &   \checkmark   &      &  \\
Maciejowska et al. 2024 \cite{maciejowska2024multiple}  &  \checkmark  & \checkmark   &  \checkmark  &    &    &    &    &  \checkmark  &  \checkmark  &     & \checkmark \\
Maciejowska et al. 2024 \cite{maciejowska2024probabilistic}  &    &   \checkmark   &   \checkmark &    &    &    &    &   & \checkmark  &  \checkmark   &   \\
Zakrzewski et al. 2024 \cite{zakrzewski2024remodels}    &    &  \checkmark  &    &    &    &    &    &    &   \checkmark  &     &  \\
Feron et al. 2024 \cite{feron2024probabilistic}         &    & \checkmark   &  \checkmark  &    &    &  \checkmark   &    &    & \checkmark   &      &  \\
Janczura et al. 2024 \cite{janczura2024expectile}       &    &  \checkmark  &    &    &    &    &    &  \checkmark  &  \checkmark  &     &  \\
Brusaferri et al. 2024 \cite{brusaferri2024line}        &    &   \checkmark  &  \checkmark    &    &    &    &    &    &    \checkmark &    &   \\
Lipiecki et al. 2024 \cite{lipiecki2024postprocessing}  &    &  \checkmark  & \checkmark   &   &  &  \checkmark  &  &  &  \checkmark  &   &   \\
Dutot et al. 2024 \cite{dutot2024adaptive}              &  \checkmark  &    & \checkmark   &    &    &    &    &    &  \checkmark   &     &   \\
Oconnor et al. 2024 \cite{o2024conformal}             &   \checkmark  & \checkmark   &  \checkmark  &    &    &    &    &    &  \checkmark  &    &   \\
Hirsch et al. 2024 \cite{hirsch2024simulation}         &    &    &    &    &    & \checkmark   &    &    &    &  \checkmark  &   \\
Pavirani et al. 2024 \cite{pavirani2024predicting}     &    &    &    &    &    &    &  \checkmark  &    &    &    &  \checkmark \\
Loizidis et al. 2024 \cite{loizidis2024electricity}    &    &    &    &    &  \checkmark  &    &    &    &  \checkmark  &    &   \\
Hirsch et al. 2024 \cite{hirsch2024multivariate}       &    &    &    &    &    &    &    & \checkmark   &    &  \checkmark  &   \\
Cornell et al. 2024 \cite{cornell2024probabilistic}    &    &  \checkmark  &    &    &    &    &    &    &    &    &  \checkmark \\
Alghumayjan et al. 2024 \cite{alghumayjan2024conformal}  &    &    &  \checkmark  &    &    &    &    &    &    &    &   \checkmark \\
Osone et al. 2025 \cite{osone2025quantile}             &  \checkmark  &    &    &    &    &    &    &    & \checkmark   &    &   \\
Lipiecki et al. 2025 \cite{lipiecki2025isotonic}       &   &  \checkmark   &  \checkmark  &    &    &    &    &    &  \checkmark  &    &   \\
Chen et al. 2025 \cite{chen2025probabilistic}          &   &    &    &    &    & \checkmark   &    &    &    & \checkmark   &   \\
Chen et al. 2025 \cite{chen2025outlier}                & \checkmark  &    &    &    &    &    &    &    &  \checkmark  &    &   \\
Yu et al. 2025 \cite{yu2025orderfusion}                & \checkmark  &    &    &    &    &    &    &    &    &  \checkmark  &   \\
Nickelsen et al. 2025 \cite{nickelsen2025bayesian}     &   &    &    &  \checkmark  &    &    &    &    &    &  \checkmark  &   \\
Pavirini et al. 2025 \cite{pavirani2025predicting}     &   &    &    &    &    &    &  \checkmark  &    &  \checkmark  &    &   \\

O'Connor et al. 2025 \cite{o2025optimising}             & \checkmark   &    &    &    &    &    &    &    &  \checkmark  &    &  \checkmark \\
Yang et al. 2025 \cite{yang2025data}             &    &    &    & \checkmark   &    &    &    &    &  \checkmark  &    &   \\
Le et al. 2025 \cite{le2025probablistic}             &    &  \checkmark  &    &    &    &    &    &    &  \checkmark  &    &   \\
O'Connor et al. 2025 \cite{o2025conformal}             &  \checkmark  & \checkmark   &  \checkmark  &    &    &    &    &    &  \checkmark  &    &  \checkmark \\ 
\end{tabular}
\caption{Literature on PEPF. Acronyms- Bayesian (BY), Bootstrap (BS), Distributional Based (DB), Monte Carlo (MC), Historical Simulation (HS),  Real-Time (RT). The \checkmark is the approach presented herein.}
\label{TableUELR}
\end{table}

Market coverage analysis shows that the DAM continues to dominate research, reflecting its centrality in electricity trading and the availability of structured price data. By contrast, the IDM and BM remain under-represented, despite their rising operational importance and the increased forecast uncertainty driven by high renewable penetration and system volatility \cite{ortner2019future}. This gap is particularly notable for modern methods such as CP, which, although increasingly used in the DAM, are rarely applied to real-time markets, despite their theoretical advantages under non-exchangeable, high-frequency conditions. While recent studies have begun to explore CP in the BM, the IDM remains largely overlooked, even as its role in managing intraday fluctuations becomes more critical. These gaps highlight a clear opportunity for methodological expansion beyond the DAM. Future work should prioritise extending adaptive PEPF methods to the IDM and BM, where participants must respond to rapid, uncertain changes in supply and demand. Techniques like SPCI and EnbPI are especially promising given their ability to dynamically recalibrate PIs while preserving empirical validity under distributional drift. At the same time, a systematic benchmarking of DL-based probabilistic models against traditional Bayesian and Monte Carlo methods could offer insights into trade-offs efficiency and validity, as well as computational cost. Hybrid strategies, particularly those combining QRA with CP, may offer a balanced path forward, combining sharpness, validity, and robustness across diverse market regimes.

\begin{table}[H]
\centering
\scriptsize 
\begin{tabular}{p{3.5cm} p{0.3cm} p{0.5cm} p{0.3cm} p{0.3cm} p{0.3cm} p{0.4cm} p{0.3cm} p{0.3cm} p{0.5cm} p{0.7cm}}
\toprule
\textbf{Reference} & \textbf{QR} & \textbf{QRA} & \textbf{CP} & \textbf{BY} & \textbf{BS} &  \textbf{DB} & \textbf{MC}  & \textbf{HS} & \textbf{PEPF} & \textbf{Other} \\
\midrule
Cottet et al. 2003 \cite{cottet2003bayesian}   &    &    &    &  \checkmark  &    &    &    &    &    &     \checkmark  \\

Clements et al. 2007 \cite{clements2007bootstrap}        &    &    &    &    &  \checkmark  &    &  \checkmark  &    &    \checkmark   \\

Do et al. 2015 \cite{do2015using}             &  \checkmark  &    &    &    &    &    &    &    &     \checkmark  & \checkmark   \\

Yong et al. 2017 \cite{yong2017neural}     &    &    &    &    &    &    &  \checkmark  &    &    &   \checkmark \\

Tibshirani et al. 2019 \cite{tibshirani2019conformal}    &    &    &  \checkmark  &  \checkmark  &    &    &    &    &  &  \checkmark   \\

Bozorg et al. 2020 \cite{bozorg2020bayesian}             &    &    &    & \checkmark   & \checkmark   &    &    &    &    & \checkmark   \\

Xu et al. 2021 \cite{xu2021conformal}                   &    &    &  \checkmark  &    &   &    &    &    &      &  \checkmark \\

Jensen et al. 2022 \cite{jensen2022ensemble}            &  \checkmark  &    &  \checkmark  &    &    &    &    &    &   \checkmark  &  \checkmark  \\
Hu et al. 2022 \cite{hu2022conformalized}               &  \checkmark   &    &  \checkmark  &    &    &    &    &    &    &   \checkmark  \\
Foygel et al. 2022 \cite{foygel2022conformal}           &    &    &  \checkmark  &   &    &    &    &    &     &   \checkmark  \\
Zaffran et al. 2022 \cite{zaffran2022adaptive}           &    &    &  \checkmark  &   &    &    &    &    &     &   \checkmark  \\

Cordier et al. 2023 \cite{cordier2023flexible}           &    &    &  \checkmark  &    &    &    &    &    &     &  \checkmark   \\
Xu et al. 2023 \cite{xu2023sequential}                   &    &    &  \checkmark  &    &    &    &    &    &    &   \checkmark  \\
Wang et al. 2023 \cite{wang2023conformal}                &    &    &  \checkmark  &    &    &    &    &    &    &  \checkmark  \\
Subhankar et al. 2023 \cite{subhankar2023probabilistically} &    &    &  \checkmark  &    &    &    &    &    &     & \checkmark   \\
Dewolf et al. 2023 \cite{dewolf2023valid}                &    &    &  \checkmark  &  \checkmark   &    &    &    &    &      &  \checkmark  \\

Xu et al. 2024 \cite{xu2024conformal}                   &    &    &  \checkmark  &    &    &    &    &    &    &  \checkmark   \\
Angelopoulos et al. 2024 \cite{angelopoulos2024conformal} &    &    &  \checkmark  &    &    &    &    &    &     &  \checkmark   \\
Lee et al. 2024 \cite{lee2024transformer}               &    &    &  \checkmark  &    &    &    &    &    &      & \checkmark  \\
Jonkers et al. 2024 \cite{jonkers2024novel}             &  \checkmark  &    &  \checkmark  &    &    &    &    &    &    &  \checkmark  \\
Renkema et al. 2024 \cite{renkema2024conformal}         &  \checkmark  &    &  \checkmark  &    &    &    &    &    &    & \checkmark   \\
\end{tabular}
\caption{Literature on probabilistic applications beyond just EPF. Acronyms- Bayesian (BY), Bootstrap (BS), Distributional Based (DB), Monte Carlo (MC), Historical Simulation (HS). The \checkmark is the approach presented herein}
\label{Table_non_UELR}
\end{table}
Table \ref{Table_non_UELR} presents a targeted overview of probabilistic forecasting methodologies applied outside core EPF. These studies span diverse energy domains, such as load, wind, demand, and PV forecasting, as well as energy storage optimisation, where uncertainty quantification is critical. Across this broader context, CP emerges as the dominant method, appearing in 16 of the 21 studies, particularly after 2019. CP’s rise reflects a wider methodological shift toward adaptive, validity-guaranteed approaches that are well-suited for high-variance, data-rich settings. Several papers, including \cite{tibshirani2019conformal} and \cite{dewolf2023valid}, combine CP with Bayesian inference, underscoring its role as a recalibration tool for improving empirical reliability. Meanwhile, traditional techniques such as Monte Carlo, bootstrap, and distribution methods have become less common. QR appears only in hybrid implementations, and QRA is notably absent, reinforcing its niche within EPF. The “Other” category, featured in 20 studies, captures the diversity of applications and hybrid modelling strategies explored in these works. Overall, this literature signals a methodological convergence around CP and ensemble-based techniques, offering compelling templates that could be adapted to EPF, particularly for under-represented, high-volatility markets like the IDM and BM.

\subsection{Cross-Market Method Comparison}
Probabilistic forecasting methods display distinct usage patterns across the DAM, IDM, and BM. The DAM remains the main target for methodological innovation, benefiting from abundant structured data and longer lead times. By contrast, the IDM and BM pose tougher challenges, scarce data, continuous trading, and extreme volatility, yet research in these markets is still sparse. This section consolidates how the main methodological classes, Bayesian and other parametric approaches, distribution, and simulation-based methods, and quantile or conformal techniques, perform across these these markets.

\subsubsection{Parametric and Non-Parametric Methods}
In the DAM, Bayesian models and distributional approaches have been widely tested. Early Bayesian inference and Monte Carlo simulation frameworks offered full predictive distributions and stress-testing capacity, but computational costs and prior sensitivity limited their uptake in real-time use. More recently, deep BNNs and hybrids with Monte Carlo have improved tail risk capture, while non-parametric techniques such as KDE, GAMLSS, and copulas have provided flexible distribution modelling  \cite{yuanchen2023electricity, nickelsen2024bayesian}. These remain useful for DAM contexts with relatively stable data availability.

In the IDM and BM, these approaches are far less established. Only a handful of IDM studies explore Bayesian DL or copula-based models, and their scalability to continuous high-frequency trading is unresolved. Distributional approaches (e.g., normalising flows, multivariate copulas \cite{hirsch2024simulation, chen2025probabilistic}) show promise for capturing evolving dependencies, but evidence is preliminary. In the BM, Bayesian and distribution-based methods have appeared sporadically, typically with modest gains over heuristics, suggesting limited practical impact without major adaptation. Overall, parametric and distributional techniques retain value in DAM but remain experimental in real-time markets.

MC and bootstrap methods follow a similar pattern. In DAM, they are common for scenario generation and ensemble calibration, often improving empirical coverage. Bootstrap methods in particular strengthen ensemble and NN forecasts under volatile conditions. Historical simulation, though simple, has also proved effective in DAM residual-based models \cite{maciejowska2024multiple, hirsch2024multivariate}. Yet in IDM/BM, evidence is fragmented: MC remains largely experimental, bootstrap shows some promise for real-time recalibration, and historical simulation is almost absent, likely reflecting its poor fit to fast-changing, non-stationary conditions.

\subsubsection{Quantile Regression, Quantile Regression Averaging, and Conformal Prediction}
Quantile-based methods have dominated DAM research since GEFCom2014. QR captures non-linearities and tail risks, while QRA provides robustness through ensemble aggregation. Despite refinements (e.g., regularisation, dimensionality reduction), QRA often struggles with calibration in volatile regimes. By contrast, CP has risen sharply since 2020, offering validity guarantees under minimal assumptions. DAM studies consistently show that CP, particularly sequential and ensemble variants such as SPCI and EnbPI, delivers more reliable coverage and, in some cases, superior economic value compared to QR and QRA \cite{o2025conformal, brusaferri2024line}. SCP remains in use but is generally too conservative for operational application.

In IDM, only isolated studies apply QR or QRA \cite{yu2025orderfusion}, typically within hybrids \cite{maciejowska2024probabilistic, kath2021conformal}. CP methods are just beginning to appear, and no studies have yet tested EnbPI or SPCI despite their potential for handling temporal drift. The BM has seen more experimentation: QR–CP hybrids combine sharpness with recalibration, and early results show SPCI outperforming both QR and QRA in validity and trading profitability. However, applications remain limited, and their scalability to ultra-low-latency settings and diverse geographical markets is largely untested.
\\
\\
\textit{Takeaways:}
\begin{itemize}
    \item \textit{DAM:} QRA remains a strong baseline, but CP hybrids now offer the best balance of sharpness and validity. Bayesian and distributional approaches still add value for stress testing.
    \item \textit{IDM:} Evidence is thin. Distributional methods (copulas, flows) are promising, but adaptive CP is notably absent, a clear research gap.
    \item \textit{BM:} Ensembles and adaptive CP, especially SPCI, shows the most consistent gains in coverage and profitability. Other methods deliver only incremental improvements, highlighting CP’s central role in this market.
\end{itemize}

\subsection{Practical Considerations}
Methodological progress in probabilistic forecasting is only valuable if it improves real-world decisions. Market participants, in particular Traders and grid operators, require forecasts that quantify uncertainty in a way that directly enhances bidding, scheduling, and risk management. This section highlights how different approaches translate into practice, with a focus on operational implications, evaluation criteria, and the persistent validity–efficiency trade-off for forecasters.

\subsubsection{Implications for Traders \& Grid Operators}
For practitioners, the key question is not which method is most novel, but which delivers reliable, actionable information under market constraints. In the DAM, Bayesian models, distributional approaches, and Monte Carlo simulations are well established, supporting scenario-based bidding and stress testing. These techniques help capture both epistemic and aleatoric uncertainty, though computational burden limits their transferability to real-time markets \cite{vahidinasab2010bayesian, yuanchen2023electricity, yang2025data}.

In the IDM and BM, evidence is thinner. Distributional methods (e.g., copulas, normalising flows) can model complex dependencies, and bootstrap resampling has shown promise for recalibrating real-time forecasts. Yet the most consistent gains for practitioners come from quantile-based models and their extensions. QR and QRA remain popular for decomposing price drivers and identifying asymmetric risks, while their DL-enhanced variants improve sharpness during spikes. Importantly, integrating QR with optimisation (e.g., Mixed-Integer Linear Programming (MILP) for storage trading) demonstrates measurable economic value \cite{o2025optimising}. 

Coverage, however, is QR’s Achilles heel, especially in high-volatility IDM and BM settings. CP offers an attractive solution, offering validity guarantees through model-agnostic recalibration \cite{dewolf2023valid, jensen2022ensemble}. SCP provides conservative intervals, while ensemble-based EnbPI and sequential SPCI deliver sharper, adaptive bounds and consistently stronger trading gains. Recent BM studies show that CP hybrids outperform QR/QRA alone in both empirical coverage and profitability \cite{o2024conformal, o2025conformal}. The lesson for practitioners is clear: DAM can rely on mature quantile and Bayesian approaches, but IDM and BM require adaptive CP to manage drift and preserve reliability in real time.

\subsubsection{Evaluation Criteria: Validity vs Efficiency}
The trade-off between validity, achieving nominal coverage, and efficiency, delivering sharp PIs, is the central tension in PEPF. In practice:
\begin{itemize}
    \item \textit{QR/QRA:} Efficient and sharp, especially in DAM, but prone to under-coverage in volatile regimes \cite{chaweewat2020electricity, xu2024novel}.
    \item \textit{CP- SCP, EnbPI, and SPCI:} Consistently valid across markets, with evidence of improved trading outcomes, though often at the cost of wider PIs \cite{o2024conformal, brusaferri2024line}.
    \item \textit{Bayesian/distributional:} Balanced validity across regimes, but heavy computation and prior sensitivity limit adoption in IDM/BM \cite{brusaferri2020probabilistic, yuanchen2023electricity}.
    \item \textit{Bootstrap/historical simulation:} Useful for recalibration and stress testing, but risk conservatism or vulnerability to non-stationarity \cite{wan2013hybrid, narajewski2022optimal}.
\end{itemize}

Recent benchmarks suggest hybrid models offer the best compromise: QR/QRA provides sharpness, while CP layers restore validity. Ensemble designs that integrate SPCI or EnbPI with QR outperform standalone methods in both statistical and financial metrics \cite{o2025conformal}. For real-world application, validity must take precedence, intervals that are sharp but unreliable undermine trust and increase imbalance costs. The operational challenge is to minimise the efficiency penalty of validity-preserving methods.
\\
\\
\textit{Takeaways:}
\begin{itemize}
    \item \textit{DAM:} Quantile-based models QR \& QRA remain effective, especially when combined with DL or CP for recalibration. Bayesian methods add value for stress testing \cite{janczura2024expectile}.
    \item \textit{IDM:} Adaptive methods are scarce. Bootstrap and distributional models show promise, but scalable CP, EnbPI and SPCI, is the missing link for reliable real-time deployment.
    \item \textit{BM:} CP hybrids, particularly SPCI \& QR, offer the most reliable coverage and strongest trading gains; standalone QR/QRA often fail under spikes \cite{o2025conformal, brusaferri2024line}.
\end{itemize}

\subsubsection{Real-Time Forecasting Challenges}
The IDM and BM present challenges that extend beyond those of the DAM. In addition to the data being non-exchangeable, real-time data with higher temporal granularity is more sensitive to temporal drift, undermining the stationarity assumptions on which many traditional models rely. Consequently, methods that perform well in the DAM often degrade when applied to real-time markets with finer temporal resolution.
QR and QRA, though dominant in DAM, show mixed results in IDM and BM. Only a handful of IDM studies demonstrate their value, typically in hybrid settings. In the BM, QR-based ensembles can capture sharpness but often fail under spikes, while QRA’s calibration issues worsen in high-frequency trading. CP offers valid, model-agnostic recalibration \cite{dewolf2023valid, kath2021conformal, cordier2023flexible}, but standard SCP struggles to produce sharp PIs in volatile BM settings, underscoring the limitations of its exchangeability assumption in temporal contexts. More advanced methods, EnbPI and SPCI, adapt better to non-stationarity by dynamically recalibrating PIs \cite{xu2024conformal, angelopoulos2024conformal}. Evidence from the Irish BM shows that SPCI delivers sharper PIs to EnbPI, stronger empirical coverage than QR and QRA, and improved trading profitability, particularly when combined with QR and EnbPI in an ensemble. Yet these methods remain underexplored in the IDM, where continuous trading and lack of data and fixed settlement points exacerbate computational and design challenges.
Complementary approaches (Bayesian inference, MC simulation, bootstrap, distributional learning) remain underdeveloped for real-time use. They can manage model uncertainty and tail risk but are limited by high computational cost or poor scalability  \cite{klein2023deep, hirsch2024simulation}. Collectively, the literature highlights the need for adaptive, low-latency methods that explicitly account for temporal drift and the breakdown of exchangeability, requirements that only the newest CP variants for time series begin to address.
\\
\\
\textit{Takeaways:}
\begin{itemize}
    \item \textit{QR/QRA:} Effective in DAM but unreliable in IDM/BM without recalibration.  
    \item \textit{SCP:} Valid but overly conservative in real time.  
    \item \textit{EnbPI/SPCI:} Best current candidates for BM; IDM remains an open frontier.  
    \item \textit{Other methods:} Bayesian, MC, and bootstrap tools show potential but lack scalable implementations for real-time deployment \cite{pavirani2024predicting, tahmasebifar2017point, serafin2022trading}.  
\end{itemize}

\subsubsection{Implications for Robust and Stochastic Optimisation}
Probabilistic forecasts do more than quantify uncertainty, they shape the optimisation problems underpinning trading, scheduling, and system balancing  \cite{9399252}. In stochastic optimisation, forecast distributions or scenarios from Bayesian, MC, or CP models are embedded into unit commitment, storage arbitrage, or bidding strategies. Performance depends directly on forecast calibration: under-dispersed intervals increase imbalance costs, while sharp but valid forecasts improve profits and reduce reserve requirements \cite{HAI2023102879}.

Robust optimisation, by contrast, relies on uncertainty sets. Here, PIs from QR, bootstrap, or CP provide natural inputs. Heuristic intervals tend to be overly conservative, while CP-based sets retain statistical validity without unnecessary risk aversion. This is especially relevant in BM contexts, where overestimating uncertainty can inflate reserve procurement.
A promising middle ground is distributionally robust optimisation (DRO). DRO defines ambiguity sets around probability distributions, and PEPF outputs (e.g., QRA or CP-calibrated distributions) can populate these sets \cite{8820049}. This alignment of statistical validity with tractable optimisation is particularly attractive for IDM and BM, where decisions must adapt to drift and high-frequency updates.
\\
\\
\textit{Takeaways:}
\begin{itemize}
    \item \textit{Stochastic optimisation:} Forecast distributions drive trading and scheduling outcomes; calibration quality is critical.  
    \item \textit{Robust optimisation:} CP-based PIs provide risk protection without the inefficiency of heuristic sets.  
    \item \textit{DRO:} An emerging bridge that uses PEPF distributions to balance validity with tractability, well suited to IDM and BM.  
\end{itemize}

\subsection{Future Research Directions}
Progress in PEPF increasingly depends on methods that adapt to the realities of real-time markets. A first priority is the development of CP techniques that remain valid under non-exchangeability and temporal drift. Recent work shows that sequential and ensemble variants (e.g., EnbPI, SPCI) can deliver sharp PIs, while retaining coverage in volatile DAM and BM conditions. Among these, SPCI is especially promising, but its application in streaming intraday environments and high-frequency balancing remains largely unexplored. Research should focus on scaling these approaches, integrating drift detection, and expanding their application beyond a handful of European markets to establish broader benchmarks.

A second priority is the systematic evaluation of economic value. Current literature demonstrates financial gains from CP-enhanced and quantile-based optimisation strategies, yet comprehensive cost–benefit analyses across markets are rare. Future studies should link forecast quality directly to trading profits, storage arbitrage, and system balancing costs. Hybrid methods, such as QR combined with CP recalibration or embedded in mixed-integer optimisation, show potential but require stress-testing under extreme volatility and fast decision cycles. Improving the economic robustness of QRA, particularly under high-frequency conditions, remains another open challenge.

Finally, reproducibility and comparability demand shared infrastructure. The field lacks standardised benchmark datasets for IDM and BM, hindering fair evaluation. Curated datasets should capture intraday price granularity, renewable forecast errors, and operational metadata such as outages or reserve activation. Coupled with common probabilistic and economic metrics, such resources would enable transparent model comparison and accelerate methodological progress.
\\
\\
\textit{Takeaways:}
\begin{itemize}
    \item \textit{Methods:} Advance adaptive CP for streaming and non-exchangeable settings; explore ensemble and drift-aware hybrids.  
    \item \textit{Economics:} Prioritise cost–benefit evaluation of forecasts in trading, storage, and balancing contexts; refine QR/QRA hybrids for high volatility.  
    \item \textit{Data:} Create open, standardised IDM/BM datasets with operational metadata to support reproducibility and benchmarking.  
\end{itemize}

In sum, the next phase of PEPF research must balance methodological rigour with operational relevance. Adaptive validity-preserving methods, rigorous economic evaluation, and shared benchmarks together form the foundation for reliable, actionable probabilistic forecasting in increasingly dynamic electricity markets.

\section{Conclusion}\label{conclusionsec}
This review has examined the evolution of PEPF methodologies, tracing the shift from early Bayesian and Monte Carlo approaches toward more flexible, data-driven techniques including QR, QRA, and CP. Among these, CP-based methods, particularly EnbPI and SPCI, stand out for their ability to deliver valid empirical coverage, adapt to non-stationarity, and improve financial performance. These capabilities reflect the growing need in electricity markets for forecasting tools that can combine statistical rigour with real-time adaptability.

While methodological advances have transformed DAM forecasting, penetration into the IDM and BM remains limited. These higher-frequency markets are becoming strategically critical due to rising renewable penetration, declining system inertia, and the demand for rapid rebalancing in response to forecast errors and unforeseen outages. In such contexts, temporal drift and non-exchangeability undermine the assumptions of many conventional approaches, making adaptive probabilistic methods such as EnbPI and SPCI, capable of dynamically recalibrating PIs under shifting distributions, particularly well-suited, yet still underutilised. Notably, in markets historically under-represented in the literature, there is growing interest in modern approaches such as DDNNs and CP, while DAM-focused research is transitioning more slowly from traditional paradigms. Closing this methodological and market gap will require not only adaptation of techniques but also more comprehensive evaluation of their economic value using metrics aligned with trading and system operations.

Future progress in PEPF will require open benchmark datasets with higher temporal granularity, capturing key intraday volatility drivers, such as renewable forecast error, reserve activation, and market coupling flows, that are essential for achieving accurate and reliable forecasts in the IDM and BM, where more detailed features are needed than in the DAM. Alongside richer datasets, advances in scalable, low-latency probabilistic models are needed so that forecasts can be integrated directly into market participant and operator decision frameworks without compromising timeliness. Equally important is improving the interpretability and trustworthiness of these models, enabling their adoption in automated trading and control systems where transparency and reliability are essential. Together, these developments can bridge the gap between methodological innovation and operational use, aligning PEPF research with the realities of increasingly dynamic, decentralised, and renewable-dominated power systems.

\section*{List of Acronyms}
\footnotesize
\begin{tabularx}{\textwidth}{lX lX}
\textbf{EPF}   & Electricity Price Forecasting
& \textbf{PEPF}  & Probabilistic Electricity Price Forecasting \\
\textbf{PI}    & Prediction Interval
& \textbf{ML}    & Machine Learning \\
\textbf{DL}    & Deep Learning
& \textbf{DAM}   & Day-Ahead Market \\
\textbf{IDM}   & Intraday Market
& \textbf{BM}    & Balancing Market \\
\textbf{TSO}   & Transmission System Operator
& \textbf{AR}    & Autoregressive \\
\textbf{ARIMA} & Autoregressive Integrated Moving Average
& \textbf{GARCH} & Generalised Autoregressive Conditional Heteroskedasticity \\
\textbf{LASSO} & Least Absolute Shrinkage and Selection Operator
& \textbf{NNs}   & Neural Networks \\
\textbf{DNN}   & Deep Neural Network
& \textbf{DDNNs} & Distributional Deep Neural Networks \\
\textbf{LSTM}  & Long Short-Term Memory
& \textbf{BNNs}  & Bayesian Neural Networks \\
\textbf{ELM}   & Extreme Learning Machines
& \textbf{QRF}   & Quantile Random Forests \\
\textbf{GEFCom} & Global Energy Forecasting Competition
& \textbf{GAMLSS} & Generalized Additive Models for Location, Scale, and Shape \\
\textbf{KDE}   & Kernel Density Estimation
& \textbf{MC}    & Monte Carlo \\
\textbf{MCMC}  & Markov Chain Monte Carlo
& \textbf{HS}    & Historical Simulation \\
\textbf{QRA}   & Quantile Regression Averaging
& \textbf{FQRA}  & Factor Quantile Regression Averaging \\
\textbf{LQRA}  & Lasso Quantile Regression Averaging
& \textbf{SQRA}  & Smoothing Quantile Regression Averaging \\
\textbf{BMS}   & Bayesian Model Averaging
& \textbf{QRM}   & Quantile Regression Machine \\
\textbf{QR}    & Quantile Regression
& \textbf{CP}    & Conformal Prediction \\
\textbf{WCP}   & Weighted Conformal Prediction
& \textbf{CQR}   & Conformalized Quantile Regression \\
\textbf{CPA}   & Conformal Prediction Averaging
& \textbf{EnCQR} & Ensemble Conformalized Quantile Regression \\
\textbf{ENBPI} & Ensemble Batch Prediction Intervals
& \textbf{SPCI}  & Sequential Predictive Conformal Inference \\
\textbf{DRO}   & Distributionally Robust Optimisation
& \textbf{MILP}  & Mixed-Integer Linear Programming \\ 

\end{tabularx}

\section*{Acknowledgments}
This work was conducted with the financial support of Science Foundation Ireland under Grant Nos. 18/CRT/6223, 16/RC/3918 and 12/RC/2289-P2 which are co-funded under the European Regional Development Fund. For the purpose of Open Access, the author has applied a CC BY public copyright licence to any Author Accepted Manuscript version arising from this submission.
\bibliography{Bibfile}

\begin{thebibliography}{100}
\bibitem{kendall2017uncertainties}
A.~Kendall, Y.~Gal, What uncertainties do we need in bayesian deep learning for computer vision?, Advances in neural information processing systems 30.
\end{thebibliography}

\end{document}